\documentclass[draft]{agujournal2019}
\usepackage{url} 
\usepackage{lineno}
\usepackage{soul}

\usepackage{calc}
\usepackage{subcaption}
\usepackage{amsmath}
\usepackage{placeins}
\usepackage{amssymb}
\DeclareMathAlphabet\mathbfcal{OMS}{cmsy}{b}{n}

\newcommand{\gt}[1]{\mathbf{\bar{#1}}} 

\let\oldalign\align
\let\oldendalign\endalign

\renewenvironment{align}
  {\linenomath\oldalign}
  {\oldendalign\endlinenomath}

\let\oldref\ref
\newcommand{\refp}[1]{(\oldref{#1})}

\draftfalse

\journalname{Water Resources Research}

\begin{document}
\title{Bayesian Poroelastic Aquifer Characterization from 
 \\ InSAR Surface Deformation Data.\\
  Part I: Maximum A  Posteriori Estimate}

\authors{Amal Alghamdi \affil{1}, Marc A. Hesse \affil{1,2}, Jingyi Chen
\affil{2,3}, Omar Ghattas \affil{1,2,4}}

\affiliation{1}{University of Texas at Austin,
Oden Institute for Computational Engineering and Sciences, 
Austin, TX, United States}
 \affiliation{2}{University of Texas
at Austin, Geological Sciences, Austin, TX, United States}
\affiliation{3}{University of Texas at Austin, Aerospace
Engineering \& Engineering Mechanics, Austin, TX, United States}
\affiliation{4}{University of Texas at Austin,
Mechanical Engineering, Austin, TX, United States}





\correspondingauthor{Marc Hesse}{mhesse@jsg.utexas.edu}



\begin{keypoints}
\item A Bayesian framework was developed for characterizing heterogeneous aquifer properties from surface deformation data.
\item Scalable algorithms allow the inference of high-dimensional discretized parameters in a transient fully coupled 3D poroelastic aquifer model.  
\item InSAR data significantly improves characterization of lateral aquifer heterogeneity at a test site in Nevada.
\end{keypoints}

%
%

%
%


\begin{abstract}


Characterizing the properties of groundwater aquifers is essential for predicting aquifer response and managing groundwater resources. In this work, we develop a high-dimensional scalable Bayesian inversion framework governed by a three-dimensional quasi-static linear poroelastic model to characterize lateral permeability variations in groundwater aquifers. We determine the maximum a posteriori (MAP) point of the posterior permeability distribution from centimeter-level surface deformation measurements obtained from Interferometric Synthetic Aperture Radar (InSAR). The scalability of our method to high parameter dimension is achieved through the use of adjoint-based derivatives, inexact Newton methods to determine the MAP point, and a Mat\'ern class sparse prior precision operator. Together, these guarantee that the MAP point is found at a cost, measured in number of forward/adjoint poroelasticity solves, that is independent of the parameter dimension. We apply our methodology to a test case for a municipal well in Mesquite, Nevada, in which InSAR and GPS surface deformation data are available. We solve problems with up to $320{,}824$ state variable degrees of freedom (DOFs) and $16{,}896$ parameter DOFs. A consistent treatment of noise level is employed so that the aquifer characterization result does not depend on the pixel spacing of surface deformation data. Our results show that the use of InSAR data significantly improves characterization of lateral aquifer heterogeneity, and the InSAR-based aquifer characterization recovers complex lateral displacement trends observed by independent daily GPS measurements.

\end{abstract}

\section{Introduction}
\label{sec: Intro}

The sustainable management of groundwater (GW) resources is quickly emerging as a critical issue as irrigated agriculture continues to grow and rely more and more on GW \cite{Hoekstra2012}. Estimates suggest that half the GW extracted for irrigation exceeds aquifer recharge and is hence unsustainable \cite{Rost2008,Vorosmarty2000}. In particular, areas with persistent water stress and large aquifer systems have increased their reliance on GW and now experience sustained GW depletion \cite{Hanasaki2008,Wada2010,ScanlonFauntLonguevergneEtAl2012}, which poses a threat to agricultural systems, food security, and livelihoods
\cite{UNESCO2012}.

Sustainable GW management under changing climatic and societal conditions requires models that can predict the aquifer response to these forcings. The dominant uncertainty in the predictions of any GW model is the insufficient characterization of the aquifer properties \cite{Eaton2006,Bohling2010,Oliver2011}. These properties can change by several orders of magnitude and display  variability on all scales. 
In principle, the uncertainty in these parameters could be quantified with Bayesian inference \cite{Carrera1986a,Carrera1986b,yeh1986,McLaughlin1996}, if a sufficiently rich data set is available. However, standard aquifer characterization based on sparse measurements at wells is often not sufficient to significantly reduce the uncertainty in model predictions \cite{Bohling2010}. Hence there is an urgent need to improve GW monitoring and integrate new data into GW models to reduce the uncertainty in predictions and to facilitate decision making.

Over the past several decades, advances in satellite remote sensing techniques have revolutionized capabilities for observing freshwater resources. In particular, spaceborne Interferometric Synthetic Aperture Radar (InSAR) has been measuring surface deformation history since 1992 with 10--100 meter spatial resolution and up to millimeter-level measurement accuracy \cite{Hanssen2001, Rosen2000}. 
The potential use of surface deformations derived from InSAR in aquifer characterization has been recognized since as early as 2000; see \citeA{galloway2007} for a review. Since then spaceborne InSAR missions such as ERS-1/2, Envisat, ALOS, and Sentinel-1 have generated large amounts of freely available data (1992--present) for monitoring the state of confined groundwater aquifers. Unlike data types that require field work and instrumentation, InSAR data can be obtained and processed at relatively low cost and effort. 

However, despite significant work on InSAR based aquifer monitoring \cite{Amelung1999, Hoffmann2001,Du2001, Lu2001,Vasco2001, Schmidt2003,galloway2007, Vasco2008,Burbey08,Bell2008, galloway2011, Chaussard2014, Chen2016, Chen2017, Miller2017}, the potential for this data for informing lateral aquifer heterogeneity has not yet been realized.
This requires scalable computational approaches that allow the inference of high-dimensional parameter fields describing the aquifer properties. It also requires the use of geomechanical models that link the observed surface deformation to GW flow in the aquifer. Such poroelastic inversion methods have only recently been developed by \citeA{IglesiasMcLaughlin12} and \citeA{HesseStadler14} and have not yet been applied to real field data. 

Here we build on the work of \citeA{HesseStadler14} and develop a high-dimensional scalable poroelastic Bayesian inversion framework to characterize lateral permeability variations in groundwater aquifers based on InSAR surface deformation data. The Bayesian framework provides a systematic and rational framework for quantifying uncertainties in the inverse solution, stemming from uncertainties in the data, model, and any prior information on the parameters, along with insensitivity of observables to parameters  \cite{Tarantola05,KaipioSomersalo05,Stuart10}. We apply this framework to a well-characterized field site in Nevada \cite{Burbey06} and focus on the inference of lateral permeability variations. In particular, we find the maximum a posteriori (MAP) point of the Bayesian posterior distribution. This application is computationally challenging and requires the inference of $16{,}896$ parameters in a three-dimensional geomechanical model with $320{,}824$ state variables. To validate our inversion results, we predict the lateral deformations using a poroelasticity model with an InSAR-inferred permeability field and compare them to available GPS data \cite{Burbey06}. In part II, we build on the Bayesian framework we present here to quantify the uncertainty in the inverse solution and resulting state variable predictions, using a tailored Markov chain Monte Carlo method.

This manuscript is organized as follows. We present the general Bayesian geomechanical inversion framework in section~\ref{sec: Method}. In section~\ref{sec: Test Case}, we describe the Nevada test case to which we apply our framework and provide details about the available InSAR and GPS data sets. Numerical results for the Nevada test site are presented in section~\ref{sec: Results} and discussed in section~\ref{sec: Discussion}. Finally, the broader conclusions of this study are presented in section~\ref{sec: Conclusion}.

\section{Methods}
\label{sec: Method}
The solution of realistic aquifer characterization problems, such as the test case in section~\ref{sec: Test Case}, requires robust, efficient, and scalable methods for both the forward and inverse problems. The Bayesian inversion framework provides a means of incorporating prior assumptions about the aquifer properties and the data noise probability distribution to infer model parameters that comply with these assumptions and with a forward model. Section~\ref{sec: Forward Model} describes the formulation and a robust discretization of the poroelastic forward problem and section~\ref{sec: Bayesian Framework} outlines the application of the Bayesian inversion framework to this forward problem with InSAR surface deformation data.

\subsection{The Forward Model}
\label{sec: Forward Model}
Here we use quasi-static linear poroelasticity to model coupled groundwater flow and elastic deformation in a confined aquifer \cite{Biot1941,showalter2000,Wang2000,Segall2010}.  We adopt a three-field formulation and a mixed discretization to conserve mass discretely  and reduce numerical oscillations \cite{Phillips2005,PhillipsWheeler2009,FerronatoCastellettoGambolati10,HagaOsnesLangtangen12}. This formulation of linear quasi-static poroelasticity in a space-time domain $\Omega\times(0,T]$ can be written as:
\begin{align}
(S_\epsilon p + \alpha \nabla\cdot\mathbf{u} )_t + \nabla\cdot\mathbf{q} =& f_p \label{equ: PDE 3 Field a}\\
-\nabla\cdot(\pmb{\sigma}(\mathbf{u}) - \alpha p \mathbf{I} ) =& \mathbf{f_u}  \label{equ: PDE 3 Field b}  \\
	\mathbf{q} + \frac{e^{m}}{\mu}\nabla p   =& 0,  \label{equ: PDE 3 Field c}
\end{align}
where $(\cdot)_t$ denotes time derivative,  $p=p(\mathbf{x},t)$ is the deviation from hydrostatic pressure, $\mathbf{q}(\mathbf{x},t)$ is the volumetric fluid flux, $\mathbf{u}=\mathbf{u}(\mathbf{x},t)$ is the displacement of the solid skeleton, $f_p(\mathbf{x},t)$ is a fluid source per unit volume, $\mathbf{f_u}(\mathbf{x},t)$ is the body force per unit volume,   $S_\epsilon$ is the specific storage, $\kappa(\mathbf{x})= e^{m(\mathbf{x})}$ is the medium permeability field, $\mu$ is the pore fluid dynamic viscosity, $\alpha$ is the Biot--Willis coupling parameter, and $\pmb{\sigma}(\mathbf{u})$ is the stress tensor. The medium permeability is written in terms of the log-permeability field $m$ to ensure that the inferred permeability field is positive. The initial and boundary conditions for the state variables $\mathbf{u}$, $p$, and $\mathbf{q}$ are given by:
\begin{align}
p(\mathbf{x}, 0) = & p_0 \text{ in } \Omega\nonumber \\
\mathbf{q}\cdot\mathbf{n} = g \text{ on } (0,T] \times \partial\Omega^n_p \quad & p = p_d  \text{ on } (0,T] \times \partial\Omega^d_p \nonumber \\
(\pmb{\sigma}(\mathbf{u})-\alpha p \mathbf{I} )\mathbf{n} = \mathbf{g} \text{ on } (0,T] \times \partial\Omega^n_{\mathbf{u}}\quad  & \mathbf{u} = \mathbf{u}_d  \text{ on } (0,T] \times \partial\Omega^d_{\mathbf{u}}, 
\end{align}
where $p_0=p_0(\mathbf{x})$ is the initial pressure at time $t=0$, $g(\mathbf{x})$ is the value of the fluid flux across the pressure Neumann boundaries $ (0,T] \times \partial\Omega^n_p$, $p_d(\mathbf{x})$ is the prescribed pressure on the pressure Dirichlet boundaries $  (0,T] \times \partial\Omega^d_p $, $\mathbf{g}(\mathbf{x})$ is the prescribed traction on the displacement Neumann boundaries $ (0,T] \times \partial\Omega^n_{\mathbf{u}} $, and   $\mathbf{u}_d(\mathbf{x})$ is the prescribed displacement on the displacement Dirichlet boundaries $ (0,T] \times \partial\Omega^d_{\mathbf{u}}$. The stress tensor in terms of displacement is given for isotropic linear elasticity by
\begin{align}
\pmb{\sigma}(\mathbf{u}) =  G (\nabla \mathbf{u} + {\nabla \mathbf{u}}^T )   +  \frac{2G\nu}{1-2\nu} (\nabla \cdot \mathbf{u}) \mathbf{I},
\end{align}
where $G$ is the drained shear modulus and $\nu$ is the Poisson's ratio.

We discretize the system~\refp{equ: PDE 3 Field a}--\refp{equ: PDE 3 Field c} with the  Mixed Finite Element Method (MFEM) proposed by \citeA{FerronatoCastellettoGambolati10}, which approximates the fluid flux in the lowest-order Raviart--Thomas space, the pressure in the space of piecewise constant functions, and the  displacement in the first-order Lagrange polynomial space. We use the implicit Euler method for integration in time. In  \oldref{app: MFEM Discretization}, we provide the details of the weak form and discretization. The resulting linear system of equations that needs to be solved at each time step is given by:
\begin{align}
	\mathbf{M} \mathbf{P}_k + \alpha \mathbf{D}\mathbf{U}_k  + {\Delta t}_k
\mathbf{D}^\mathbf{q}\mathbf{Q}_k =&{\Delta t}_k \mathbf{F}^p_k +\mathbf{M} \mathbf{P}_{k-1} + \alpha
\mathbf{D}\mathbf{U}_{k-1} \nonumber \\
	- \alpha\mathbf{G}\mathbf{P}_k - \mathbf{E}\mathbf{U}_k   =&-
	  \mathbf{F}^\mathbf{u}_k - \mathbf{F}^{\mathbf{u}, bcs}_k
 \nonumber  \\
        {\Delta t}_k \mathbf{G}^p  \mathbf{P}_k -  {\Delta t}_k \mathbf{K}
\mathbf{Q}_k =&   {\Delta t}_k \mathbf{F}^{p,bcs}_k, \label{equ: Discretized 3 Field System} 
\end{align}
where $\mathbf{P}_k,\; \mathbf{U}_k$ and $\mathbf{Q}_k$ are the discretized pressure, displacement and fluid flux degrees of freedom (DOFs) vectors, respectively. The subscript $k$ denotes the evaluation of the variable (or the operator) at time $t_k$ and  ${\Delta t}_k= t_k-t_{k-1}$ is the $k$-th time step size. The discrete differential operators are denoted $\mathbf{D}$ and  $\mathbf{D}^\mathbf{q}$ for the divergence of the displacement and the flux, $\mathbf{G}$ and $\mathbf{G}^p$ for the gradient of the pressure in the pressure and the mixed flow spaces, respectively, and $ \mathbf{E}$ for the linear elastic operator. The mass matrices $\mathbf{M}$ and $ \mathbf{K} $ are  weighted by the specific storage and mobility (permeability/viscosity), respectively. Right hand side contributions include the discrete source terms for fluid and body force, $ \mathbf{F}^p_k$ and $\mathbf{F}^\mathbf{u}_k $, as well as boundary tractions, $\mathbf{F}^{\mathbf{u}, bcs}_k$, and pressure natural boundary conditions,  $\mathbf{F}^{p,bcs}_k$.

To facilitate the derivation of the time-dependent inverse problem in  section~\ref{sec: MAP}, we follow \citeA{HesseStadler14} and define the ``global'' space-time system that represents the solution for all time steps simultaneously and denote it as
\begin{align}
	\gt{S}\gt{X} = \gt{F}, \label{equ: Global Time Space System}
\end{align}
where the block lower triangular matrix $\gt{S}$ combines the differential operators of system~\refp{equ: Discretized 3 Field System}, $\gt{F}$ combines the source terms and the natural boundary conditions, and $\gt{X}$ consists of the pressure, displacement, and fluid flux DOFs at all time steps (see \oldref{app: MFEM Discretization} for the explicit expressions).  The bar notation is used to distinguish the space-time discretized operators and vectors from the space-only discretized operators and vectors.

\subsection{Bayesian Framework for Three-Field Formulation of  Fully Coupled 3D Poroelasticity Inverse Problem }
\label{sec: Bayesian Framework}
The Bayesian inverse problem solution is a probabilistic characterization of unknown physical system parameters, such as material properties, source terms, boundary conditions, and initial conditions. Bayes' theory provides a framework for updating ``prior'' statistical knowledge about these parameters by using observational data and a mathematical model of the system via a ``likelihood distribution''. 
This likelihood distribution determines how likely it is that the observed data result from a particular parameter realization. The updated probability distribution is known as the posterior distribution, and is regarded as the solution of the inverse problem. Bayesian inversion, therefore, provides a characterization of the uncertainty in the parameters due to uncertainty in the prior information, data, and model, as opposed to finding a ``point estimate'', as is the case with deterministic inversion. In the present article (Part I), we formulate the Bayesian inversion problem, paying special attention to the prior distribution of the model parameters and the observational noise model. We also take the first step in exploring the posterior distribution by introducing fast Hessian-based methods for scalably determining the maximum a posteriori (MAP) point. In a subsequent article (Part II), we will build on the tools developed here to construct a Hessian-driven Markov chain Monte Carlo method (MCMC) to sample the posterior distribution and compute statistics of interest as well as posterior predictives.

In section~\ref{sec: Bayesian Inversion}, we describe the formulation and discretization of  the Bayesian inverse problem for  the three-field formulation of the linear quasi-static poroelasticity model described in section~\ref{sec: Forward Model}. Our methodology follows closely the framework presented by \citeA{HesseStadler14}, with the following differences or extensions: (1) we base the framework on the three-field Biot system formulation rather than the classic two-field, (2)  we use a Mat\'ern class prior for the Bayesian inverse problem (section~\ref{sec: Prior Matern}), and (3) we define the likelihood distribution based on a noise model for InSAR surface deformation data (section~\ref{sec: Likelihood}).

\subsubsection{Formulation of the Infinite Dimensional Bayesian Inverse Problem and its Discretization}
\label{sec: Bayesian Inversion}
  
Given the observational data, $\mathbf{d}^{\text{obs}}  \in {\rm I\!R}^{n_\text{obs}}$, and prior statistical knowledge about a parameter field $m$, we seek the posterior probability distribution of $m$. The parameter $m$ is related to the observational data through a Gaussian additive noise model,
\begin{align}
\mathbf{d}^{\text{obs}} = \mathcal{F}(m) + \eta \label{equ: Additive Noise Form}.
\end{align}
The \textit{parameter-to-observable map}, $\mathcal{F}(\cdot):  \mathcal{Z} \rightarrow  {\rm I\!R}^{n_{\text{obs}}}  $,  maps a realization of the parameter field from the infinite dimensional space $\mathcal{Z}$ to the finite dimensional space of observables $ {\rm I\!R}^{n_{\text{obs}} }$. Evaluating $\mathcal{F}(m)$ involves solution of a system of PDEs in which the parameter $m$ is a coefficient, source term, boundary condition, or initial condition of the system, or a combination thereof. In our case, we will invert for the log-permeability field $m$ appearing in equation~\refp{equ: PDE 3 Field c}. The parameter-to-observable map $\mathcal{F}(m)$ involves solution of the poroelasticity model~\refp{equ: PDE 3 Field a}--\refp{equ: PDE 3 Field c}, and extraction of surface displacements.   

The random variable $\eta$ is observational noise with known statistical properties. Here, $\eta$ is assumed to follow a Gaussian distribution with mean zero and covariance matrix $ \mathbf{\Gamma}_\text{noise} \in  {\rm I\!R}^{n_{\text{obs}} \times n_{\text{obs}} }$, reflecting the estimated noise level and correlation. The likelihood probability measure, which is the probability measure of the difference between the observational data and the simulated observations, $ \mathbf{d}^{\text{obs}} - \mathcal{F}(m)$, follows the normal distribution of the observational data noise.

A general framework for infinite dimensional Bayesian inverse problems governed by PDEs is presented in \citeA{Stuart10}. In infinite dimensions, Bayes' formula reads
\begin{align}
\frac{d\mu_\text{post}}{d\mu_\text{prior}} = \frac{1}{Z}\pi_\text{likelihood}(\mathbf{d}^\text{obs} | m),
\end{align}
where $\pi_\text{likelihood}$ is the likelihood probability density function,  $Z = \int_\Omega \pi_\text{likelihood}(\mathbf{d}^\text{obs} | m) d\mu_\text{prior}$ is the normalization constant required for the posterior measure, $\mu_\text{post}$, to be a probability measure, and $\frac{d\mu_\text{post}}{d\mu_\text{prior}} $ is the Radon--Nikodym derivative of the posterior with respect to the prior. 

As a first step to characterising the posterior $\mu_\text{post}$, we wish to find the point that maximizes the posterior, the so-called MAP point. We follow the \textit{discretize-then-optimize} approach because it guarantees a consistent discretized gradient \cite{Gunzburger03}. The corresponding Bayes formula in finite dimensions is given by:
\begin{align}
\pi_\text{post}(\mathbf{m}| \mathbf{d}^\text{obs}) \propto \pi_\text{likelihood}(\mathbf{d}^\text{obs} | \mathbf{m}) \pi_\text{prior}(\mathbf{m}), \label{equ: Bayes Formula Finite}
\end{align} 
where the symbol $\propto$ indicates equality up to a constant, $\mathbf{m}\in {\rm I\!R}^{n_n}$ denotes the discretized parameters,
\begin{align}
 \pi_\text{prior}(\mathbf{m}) =&
\mathcal{N}(\bar{\mathbf{m}},\mathbf{\Gamma}_\text{prior}), \quad \text{and} \quad
  \pi_\text{likelihood}(\mathbf{d}^\text{obs} | \mathbf{m})  \propto  e^{-\ell(
 \mathbf{d}^\text{obs};\mathbf{m})}. 
\end{align} 
Here $\mathcal{N}$ is the normal distribution of the prior with mean $\bar{\mathbf{m}} \in  {\rm I\!R}^{n_n}$ and covariance matrix $\mathbf{\Gamma}_\text{prior} \in  {\rm I\!R}^{n_n\times n_n }$.
The function $\ell$, the negative log of the likelihood, is obtained from equation~\refp{equ: Additive Noise Form} and the fact that $\eta \sim \mathcal{N}(0, \mathbf{\Gamma}_\text{noise})$ as follows:
\begin{align}
\ell( \mathbf{d}^\text{obs}; \mathbf{m}) = \frac{1}{2}(  \mathbfcal{F}(\mathbf{m}) -
\mathbf{d}^{\text{obs}})^T\mathbf{\Gamma}_\text{noise}^{-1}(   \mathbfcal{F}(\mathbf{m}) -
\mathbf{d}^{\text{obs}} ),
\label{equ: Negative Log Liklihood}
\end{align}
where $\mathbfcal{F}$ is the discretized parameter-to-observable map. The negative log of the likelihood, $\ell$, is also known as the ``data misfit term" in the deterministic inversion setting. A typical choice of $ \mathbf{\Gamma}_\text{noise} $ is a diagonal matrix with entries  $ \sigma_i^2$, the noise-level variance of the $i^{th}$ observation $\mathbf{d}^\text{obs}_i$. This choice of noise is valid for cases in which the noise polluting one measurement is uncorrelated with the noise in other measurements.

\subsubsection{Prior}
\label{sec: Prior Matern}
As a prior, we choose a Mat\'ern class that can represent several commonly used covariance models used in geostatistics \cite{Goovaerts1997,Deutsch1998,Wackernagel2010}.
Our representation of the Mat\'ern prior is based on the link between the Mat\'ern class of Gaussian fields and solutions of elliptic stochastic partial differential equations (SPDE) \cite{LindgrenRueLindstroem11}. We define the parameter $m(\mathbf{x})$ to be a Mat\'ern class Gaussian random field in the domain $\Omega \in {\rm I\!R}^d$, $d=1$, $2$, or $3$. \citeA{LindgrenRueLindstroem11} show that its covariance operator is the square of the solution operator of the SPDE
\begin{align}
(\delta - \gamma \Delta )^\frac{\alpha}{2} m(\mathbf{x})  &= s(\mathbf{x}) \;\; \text{in} \;\; \Omega.\label{equ: Prior Elliptic SPDE} 
\end{align}
For $d= 3$, $\alpha$  should satisfy $\alpha > 3/2$, we choose $\alpha = 2$.
The right-hand side of the SPDE~\refp{equ: Prior Elliptic SPDE}, $s(\mathbf{x})$, is a white-noise Gaussian random field with unit marginal variance. The parameters $\gamma$, $\delta$, and $\alpha$ are chosen to achieve the desired marginal variance and correlation length of the features of the parameter field. 

The parameter $m$ is discretized over the same triangularization defined for the linear poroelasticity PDE problem. We use first-order Lagrange polynomials to approximate $m$. For our choice $\alpha = 2$, the Mat\'ern class precision operator, $\mathbf{R}$, is constructed as follows: $\mathbf{R} = \mathbf{A}\mathbf{M}_{m}^{-1}\mathbf{A}$. The operator $\mathbf{A}$ is the finite element discretization of the differential operator $\mathcal{A} = (\delta - \gamma\Delta )$ and $\mathbf{M}_{m}$ is the mass matrix in the finite element space where $m$ is approximated.  $\mathbf{M}_{m}^{1/2}$  can be efficiently approximated  \cite{VillaPetraGhattas2019}, in which case the cost of sampling from the prior is mainly the cost of solving the elliptic PDE~\refp{equ: Prior Elliptic SPDE} for a realization of the white noise right hand side, which can be carried out with multigrid solvers which have linear complexity. The covariance of the prior, $\mathbf{\Gamma}_\text{prior}$, is given by $\mathbf{R}^{-1}$, the inverse of the precision operator. 

For an isotropic Gaussian field of Mat\'ern class, the correlation length (or range) is $\rho = \sqrt{\left(8\gamma(\alpha - \frac{d}{2})\right)/\delta}$; in particular, $m(\mathbf{x}_1)$ and $m(\mathbf{x}_2)$, for any $\mathbf{x}_1$, $\mathbf{x}_2 \in \Omega$ of distance $\rho$ from each other, have approximately a correlation near $0.1$ \cite{LindgrenRueLindstroem11}. In all results presented here, the domain is three-dimensional ($d=3$) and we choose the elliptic operator power $\alpha/2 = 1$, so that  $\rho$ reduces to $ 2\sqrt{\gamma/\delta}$. Choosing  $\frac{\gamma}{\delta} =10^{6}$, for example,  dictates a correlation near 0.1 for points that are about $2{,}000$ length units apart.

\subsubsection{Likelihood and Data Noise of InSAR Data}
\label{sec: Likelihood}

 Interferometric SAR  (InSAR) computes the phase difference between two synthetic aperture radar (SAR) images. The resulting interferogram can be used to infer a 2D map of surface deformation between two SAR acquisition times along the radar line-of-sight (LOS) direction. The LOS deformation, $u_\text{LOS}$, can be described by the formula: 
\begin{align}
u_\text{LOS}= \alpha_1 u_1 +  \alpha_2 u_2 +  \alpha_3 u_3,
\label{equ:los}
\end{align}
where $u_1$, $u_2$ and $u_3$ are the east, north and vertical displacements, accordingly. The vector $\alpha = [\alpha_1, \alpha_2, \alpha_3]$ is the radar LOS direction unit vector that can be calculated based on the known radar imaging geometry.

We define the pointwise InSAR data misfit function (i.e., the negative log likelihood), $\ell^\text{InSAR}$, as follows:
\begin{align}
\ell^\text{InSAR}(\mathbf{m}) = \sum_{i=1}^{n_\text{interfero}} 
                       \frac{1}{2(\sigma_\text{InSAR}^i)^2}
                       \sum_{j=1}^{n^i_\text{pixels}}
                       \Big( u_{\text{LOS},j}^i -u_{\text{InSAR},j}^i \Big)^2,
\label{equ: InSAR Misfit}
\end{align}
where $n_\text{interfero}$ is the number of interferograms used in the inversion and $n^i_\text{pixels}$ is the number of pixels in the $i^\text{th}$ interferogram. In this data misfit expression, we assume the covariance operator of the noise is diagonal, where $\sigma_\text{InSAR}^i$ is the noise level of the $i^{th}$ interferogram. This assumption is valid when InSAR measurement noise mostly stems from phase decorrelation as in the Nevada test case, presented in section~\ref{sec: Test Case}. In cases where InSAR measurement noise contains a non-negligible spatially coherent component (e.g.\ noise from atmospheric delays), the covariance operator can be constructed to dictate that noise correlation. The value $u_{\text{InSAR},j}^i$ is the LOS deformation measured from the $j^\text{th}$ pixel in the $i^\text{th}$ interferogram, and $u_{\text{LOS},j}^i$ is the simulated LOS deformation of the $j^\text{th}$ pixel in the $i^\text{th}$ interferogram as computed from the forward poroelasticity model.

\subsubsection{Finding the MAP Point Estimate of the Posterior Distribution}
\label{sec: MAP}

The discretized posterior~\refp{equ: Bayes Formula Finite} in general is a distribution in high dimensional parameter space---thousands or millions. It is difficult to explore such distributions due to the high dimensionality and the need to solve the 3D poroelasticity system~\refp{equ: PDE 3 Field a}--\refp{equ: PDE 3 Field c} to evaluate the posterior at each point in parameter space. Here in Part I, we seek to determine the MAP point, which as we will see, amounts to a deterministic inverse problem with specially-chosen weighting matrices. We employ adjoint-based derivatives, the Gauss--Newton method, and early termination of conjugate gradient (CG) iterations at each Gauss--Newton step to ensure the efficiency and scalability to high parameter dimensions for solving this deterministic inverse problem.

The MAP point, $\mathbf{m}_\text{MAP}$, is the point that maximizes the posterior distribution~\refp{equ: Bayes Formula Finite}. We can define the parameter-to-observable map, $\mathbfcal{F}(\mathbf{m})$, as follows:   $\mathbfcal{F}(\mathbf{m}):= \gt{B}\gt{X}$, where the state variables $\gt{X}$ depend on the permeability parameter vector $\mathbf{m}$ through the forward problem $\gt{S}\gt{X} = \gt{F}$ \refp{equ: Global Time Space System}, and $\gt{B}$  is the linear observation operator that extracts the observations from $\gt{X}$, i.e.\ $\gt{B}\gt{X}$ is the surface displacements at times and locations of the observed data. Thus we can rewrite the the posterior distribution~\refp{equ: Bayes Formula Finite} as:
\begin{align}
\pi_\text{post}(&\mathbf{m}| \mathbf{d}^\text{obs}) \propto \nonumber\\
  \text{exp}&\left(- \frac{1}{2}(
\gt{B}\gt{X} - \mathbf{d}^{\text{obs}}
)^T\mathbf{\Gamma}_\text{noise}^{-1} ( \gt{B}\gt{X} -
\mathbf{d}^{\text{obs}}  )  -  \frac{1}{2}(\mathbf{m} -
\bar{\mathbf{m}})^T\mathbf{\Gamma}_\text{prior}^{-1} (\mathbf{m} -
\bar{\mathbf{m}})\right). 
\end{align}
Maximizing the posterior distribution with respect to the parameter $\mathbf{m}$ is equivalent to minimizing the negative log posterior. Therefore, our goal is to minimize the objective function:
\begin{align}
\min_{\mathbf{m}\in {\rm I\!R}^{n_n}} \mathcal{J}(\mathbf{m}) = 
\frac{1}{2}(\gt{B}\gt{X} - &\mathbf{d}^{\text{obs}}  )^T
\mathbf{\Gamma}_\text{noise}^{-1}( \gt{B}\gt{X} -\mathbf{d}^{\text{obs}})
 +  \frac{1}{2}(\mathbf{m} - \bar{\mathbf{m}})^T\mathbf{R}
 (\mathbf{m} - \bar{\mathbf{m}}),
 \label{equ: Constrained Minimization}
\end{align}
where $\gt{X}$ depends on $\mathbf{m}$ through the solution of the forward problem $\gt{S}(\mathbf{m})\gt{X} = \gt{F}$. It can be seen that $  \mathbf{\Gamma}_\text {prior}^{-1}=\mathbf{R} $ acts as a regularization operator in the deterministic problem~\refp{equ: Constrained Minimization}\cite{Tarantola05}. To derive the gradient using the adjoint method, we construct a Lagrangian by adding to the objective function $\mathcal{J}(\mathbf{m})$ the inner product of the left-hand side of the state equation $\gt{S}(\mathbf{m})\gt{X} - \gt{F} = 0$ with a Lagrange multiplier (also called the adjoint variable)  $\gt{Y}$. In the Lagrangian formulation, $\gt{X}$ is considered an independent variable, and the dependence of $\gt{X}$ on the parameters $\mathbf{m}$ is enforced through the Lagrange multiplier term. 

At a minimum of \refp{equ: Constrained Minimization}, the partial derivatives of the Lagrangian with respect to the state variable, ${\partial\mathcal{L}}/{\partial \gt{X}}$, the parameter, ${\partial\mathcal{L}}/{\partial \mathbf{m}}$, and  the adjoint variable, ${\partial\mathcal{L}}/{\partial \gt{Y}}$, vanish. We set  ${\partial\mathcal{L}}/{\partial \gt{X}}$ and  ${\partial\mathcal{L}}/{\partial \gt{Y}}$ to zero, which gives the state equation and the adjoint equation respectively. We satisfy these two conditions by solving the discretized state and adjoint equations exactly for any value of $\mathbf{m}$. Thus we seek the parameter $\mathbf{m}$ that ensures that the gradient  ${\partial\mathcal{L}}/{\partial \mathbf{m}}$ vanishes. We solve the equation ${\partial\mathcal{L}}/{\partial \mathbf{m}} =0$ using Newton method. The details of forming the Lagrangian and constructing the Newton iteration are left to \oldref{app: Newton Derivation}.

The Newton system we need to solve at each iteration  $l$ is:
\begin{align}
 \mathbf{H}^l\delta\mathbf{m}^{l} = -\mathbf{G}^l,
 \label{equ: Newton Iteration}
\end{align}
where $\mathbf{G}^l=\mathbf{G}(\mathbf{m}^l) =  {({\partial\mathcal{L}}/{\partial \gt{X}})}^l$ is the gradient and $\delta\mathbf{m}^{l}$ is the Newton direction. The superscript $l$ we use for the matrices, and the forward and the adjoint solutions  indicates that these matrices and vectors are evaluated at the $l^\text{th}$ iteration MAP point approximation, $\mathbf{m}^{l}$. The Hessian $ \mathbf{H}^l= \mathbf{H}(\mathbf{m}^l) \in  {\rm I\!R}^{n_n \times n_n}$ is given by
\begin{align}
 \mathbf{H} = &
 \mathbf{R} + \mathbf{R}_{\mathbf{m} \mathbf{m}} +\gt{C}^T \gt{S}^{-T} (
\mathbf{W}_{\gt{X}\gt{X}} \gt{S}^{-1} \gt{C} -   \mathbf{W}_{\gt{X}\mathbf{m}} )-
\mathbf{W}_{\mathbf{m} \gt{X}}\gt{S}^{-1}\gt{C}.
\label{equ: Full Hessian}
\end{align}
The operator $\gt{C}$ is the partial derivative (sensitivity) of the residual $\mathcal{R} = \gt{S}\gt{X} - \gt{F}$ with respect to the parameter $\mathbf{m}$
\begin{align}
\gt{C} = \frac{\partial }{\partial \mathbf{m}} (\gt{S}\gt{X} - \gt{F}) =
\frac{\partial }{\partial \mathbf{m}} (\gt{S}\gt{X} ),
\end{align}
and the other matrices are defined as follows 
\begin{align}
\mathbf{W}_{\gt{X}\gt{X}} = \gt{B}^T\mathbf{\Gamma}_\text{noise}^{-1}\gt{B},\ \mathbf{W}_{\gt{X}\mathbf{m}}= \frac{\partial}{\partial\mathbf{m}}(\gt{S}^{T}\gt{Y}),\ \mathbf{W}_{\mathbf{m} \gt{X}} = \frac{\partial}{\partial \gt{X}}(\gt{C}^{T}\gt{Y}),\ \mathbf{R}_{\mathbf{m} \mathbf{m}}=   \frac{\partial}{\partial \mathbf{m}} (\gt{C}^{T}\gt{Y}).\nonumber
\end{align}

The direction $\delta\mathbf{m}^{l}$ is used to update the approximation of the MAP point $  \mathbf{m}^{l}$ as follows:
\begin{align}
\mathbf{m}^{l+1} = \mathbf{m}^{l} + \theta \delta\mathbf{m}^l,
\end{align} 
where $\theta$ is a step length parameter chosen to satisfy the Armijo rule, to guarantee a sufficient reduction in the objective function \cite{NocedalWright06}. Backtracking line search is used to reduce the step size until the objective function is sufficiently reduced. The conjugate gradient method with early termination is used to solve the linear system~\refp{equ: Newton Iteration} at each Newton iteration. This CG iteration is also terminated when a direction of  negative curvature is detected, which ensures that the direction $\delta\mathbf{m}^l$ is a  descent direction \cite{NocedalWright06}.

The full Hessian~\refp{equ: Full Hessian} can be approximated by the Gauss--Newton Hessian $\mathbf{H}_\text{GN}$, in which all matrices depending on the adjoint variable $\gt{Y}$, i.e.\ $\mathbf{W}_{\gt{X}\mathbf{m}}$, $\mathbf{W}_{\mathbf{m} \gt{X}}$, and $\mathbf{R}_{\mathbf{m} \mathbf{m}}$, are neglected. The reasoning behind this is that, when the data misfit $\gt{B}\gt{X} - \mathbf{d}^{\text{obs}} $ is zero (i.e.\ when the model fits the data), the right hand side of the adjoint equation~\refp{equ: Adjoint Equation} vanishes, and since the equation is linear in the adjoint variable $\gt{Y}$, it vanishes. Thus when the data misfit is small, the Gauss--Newton approximation can be expected to be accurate. The resulting approximation is guaranteed to be positive definite for appropriately chosen regularization, is computationally less expensive, and is expected to converge super linearly near the optimal point, $\mathbf{m}_\text{MAP}$, in the zero-noise case. However, if the noise level is high, we can switch to the full Newton Hessian after taking some initial iterations with $\mathbf{H}_\text{GN}^l $. The GN Hessian is given by
\begin{align}
\mathbf{H}_\text{GN}^l = \mathbf{R} +\gt{C}^{l^T} \gt{S}^{l^{-T}} ( \mathbf{W}_{\gt{X}\gt{X}} \gt{S}^{l^{-1}} \gt{C}^l).
\end{align}

Forming the GN Hessian (and similarly the full Hessian) explicitly  each Newton iteration $l$ \refp{equ: Newton Iteration} is computationally intractable. It requires applying the operator $\gt{S}^{l^{-1}}$ to each column of  $\gt{C}^l$ to form the matrices product $\gt{S}^{l^{-1}} \gt{C}^l$, each Newton iteration $l$. Applying the operator $\gt{S}^{l^{-1}}$ to a vector amounts to an incremental forward poroelasticity solve, equation~\refp{equ: Incremental Forward}. Therefore, forming the GN Hessian explicitly costs $n_n$ poroelasticity PDEs solves (note that no additional solves are required for forming the product $\gt{C}^{l^T} \gt{S}^{l^{-T}}$ since it is the transpose of the product $\gt{S}^{l^{-1}} \gt{C}^l$).  When using CG, there is no need to form the GN Hessian in each Newton iteration $l$ explicitly. What is required, in each CG iteration $i$, is  computing the matrix vector product $\mathbf{R}^{-1}\mathbf{H}_\text{GN}^l\mathbf{d}_{i}$, where $\mathbf{d}_{i}$ is the $i^\text{th}$ conjugate gradient direction and $\mathbf{R}^{-1}$ is the prior covariance matrix which is commonly a favorable choice as a preconditioner for the Hessian. The main cost of computing this product is the cost of applying the operators $\gt{S}^{l^{-1}}$ and $\gt{S}^{l^{-T}}$ each to a vector wich is the cost of two poroelasticity PDEs solves (one incremental forward~\refp{equ: Incremental Forward} and one incremental adjoint~\refp{equ: Incremental Adjoint}) each CG iteration. We emphasize that it is often the case that the number of CG iterations $n_\text{CG}$ required in each (Gauss--)Newton iteration $l$ is much smaller than, and independent of, the number of parameters $n_n$ ($n_\text{CG} \ll n_n$) in many PDE-based optimization problems including those governed by poroelasticity, as has been proven for some particular problems and observed numerically for a number of others \cite{BashirWillcoxGhattasEtAl08, FlathWilcoxAkcelikEtAl11, Bui-ThanhGhattas12a, Bui-ThanhGhattas13a, Bui-ThanhGhattas12, Bui-ThanhBursteddeGhattasEtAl12_gbfinalist, Bui-ThanhGhattasMartinEtAl13,ChenVillaGhattas17, AlexanderianPetraStadlerEtAl16, AlexanderianPetraStadlerEtAl17, AlexanderianPetraStadlerEtAl14, CrestelAlexanderianStadlerEtAl17, PetraMartinStadlerEtAl14, IsaacPetraStadlerEtAl15, MartinWilcoxBursteddeEtAl12, Bui-ThanhGhattas15, HesseStadler14}. As a consequence of ill-posedness, a typically small and parameter dimension-independent number of CG iterations is required, corresponding to the dominant eigenvalues of the prior preconditioned (Gauss--)Newton Hessian that are distinct from those that cluster around one. For sufficiently large CG tolerance, the eigenvector components of the error corresponding to the remaining eigenvalues clustered around one can be eliminated in $O(1)$ CG iteration \cite{NocedalWright06}. Furthermore, in early Newton iterations when the MAP point approximation $\mathbf{m}^l$  is away from the true solution, solving the linear system arising in the Newton iteration $l$ requires even fewer CG iterations than the number of the dominant eigenvalues $\sim n_\text{CG}$ due to imposing  Eisenstat-Walker tolerance that prevents oversolving the linear system   \cite{EisenstatWalker96,NocedalWright06,VillaPetraGhattas2019}.

\section{Nevada Test Case}
\label{sec: Test Case}

\begin{figure}[htp]
\includegraphics[width=1.\linewidth]{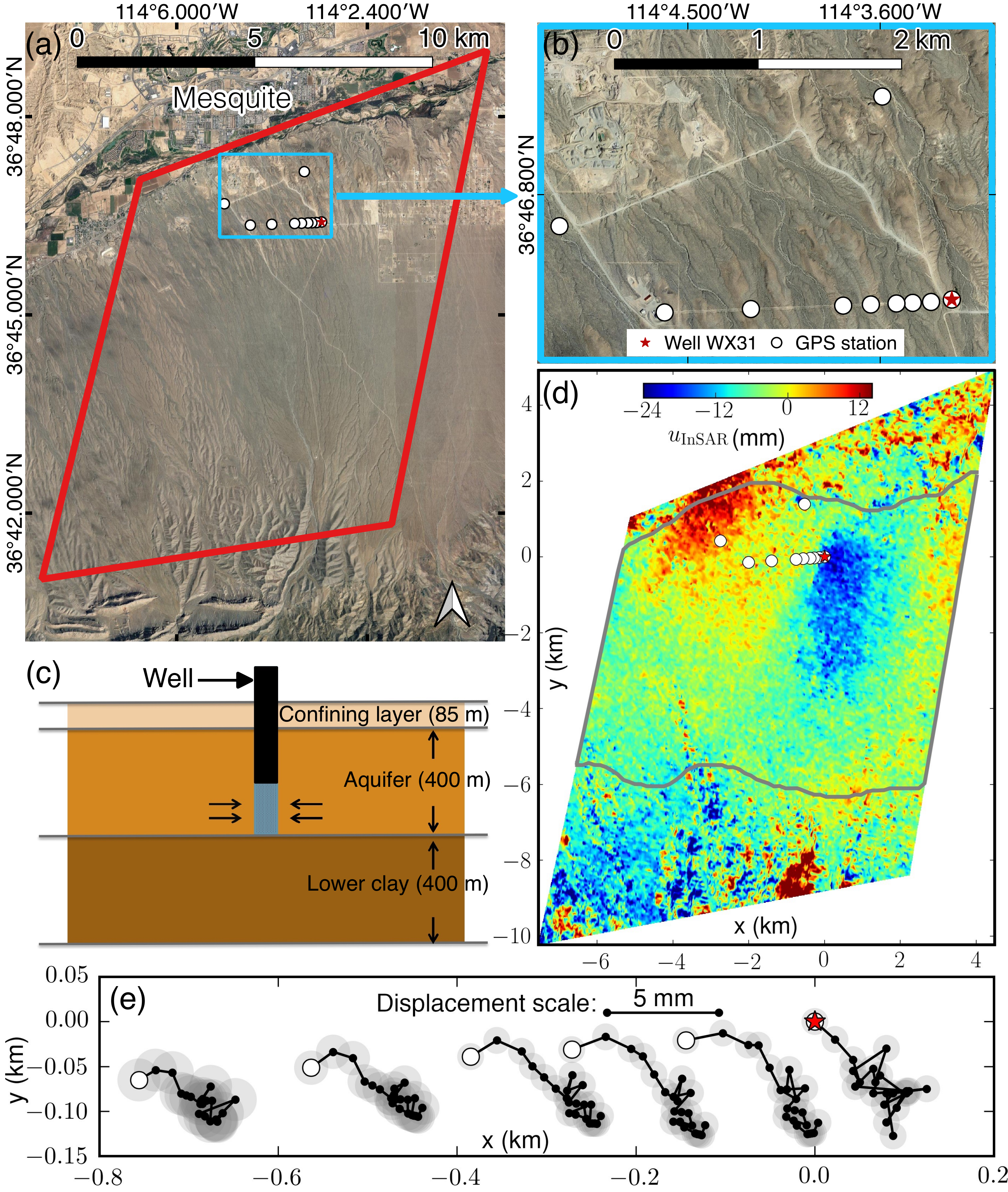}
\caption{(a) The study site near Mesquite, NV. 
         The red polygon shows the lateral boundaries of our numerical simulation domain.
         (b) Zoom-in view of the area outlined in cyan in Figure~\ref{fig: Test Case}\emph{a}.
         The red star marks the location of well WX31, where \protect\citeA{BurbeyWarnerBlewittEtAl06} performed the aquifer test in 2003. Surface deformation was monitored daily using a network of 10 high-precision GPS stations (white dots).
         (c) A conceptual illustration of the aquifer model. The screened segment of the well is marked in blue.
         (d) InSAR-observed LOS deformation between May 4, 2003 and October 26, 2003 over the study area. The the gray polygon marks the extent of the data used in the inversion.
         (e) A map view of the observed cumulative lateral displacements at 6 GPS stations that are closer to the well from day $1$ to day $22$ of the pumping test. Each black dot shows the daily cumulative lateral displacement. The gray shaded circles enclosing the GPS data are the $99\%$ confidence ellipses of the measured data. In both Figure~\ref{fig: Test Case}\emph{d} and \ref{fig: Test Case}\emph{e}, $x$ is the east direction and $y$ is the north direction relative to the well location (x,y)=(0,0).
         }
\label{fig: Test Case}
\end{figure}

In this section, we first review an aquifer pumping test conducted in 2003 by \citeA{BurbeyWarnerBlewittEtAl06} near Mesquite, Nevada. We then introduce the InSAR and GPS data used to measure the aquifer deformation induced by this test. We finally describe the aquifer model, and how we set up the Bayesian inversion framework to infer the lateral permeability variations of this aquifer.

\subsection{The Study Site}
\label{sec: Nevada Site}

The Nevada study site is located southeast of Mesquite, Nevada, in the Mesquite basin of the lower Virgin River valley (Figure~\ref{fig: Test Case}\emph{a} and \ref{fig: Test Case}\emph{b}). The aquifer is $\sim$ $400$ meter thick, which is confined by a brittle unsaturated layer at the top and a clay layer at the bottom (Figure~\ref{fig: Test Case}\emph{c}). Hydraulic head records suggest that the regional GW recharge is due mainly to winter precipitation in the surrounding mountains, and hence is negligible during the duration of the well test \cite{BurbeyWarnerBlewittEtAl06,Burbey08}. 

A controlled aquifer test was performed at a newly installed municipal well WX31 between May 7, 2003 and July 9, 2003 \cite{Burbey06}. The well was screened over a $237.7$ meter interval in the deeper portion of the confined aquifer, and produced an average of $9{,}028$ cubic meters of water daily. The pumping continued until $\sim$ October, 2003. Note that the aquifer was isolated from poromechanical stresses exerted by pumping activities in adjacent wells. Over the test period and within a radius of $\sim 9$ km from well WX31, only a single well, located 4 km toward the north-northwest of well WX31, was in production and its radius of influence did not intersect with that of WX31 \cite{BurbeyWarnerBlewittEtAl06}.

Surface deformation caused by pumping was measured daily at 10 high-precision GPS stations between April 30, 2003 and July 9, 2003. Using these measurements, \citeA{Burbey08} estimated the mean hydraulic conductivity, Poisson's ratio, and shear modulus of the aquifer. They further postulated the existence of a fault $\sim$ 1.1 km northwest of the well based on an EnviSAT interferogram that spans May 4, 2003 and September 5, 2004.

The relatively simple geologic setting along with the availability of the GPS and InSAR deformation data make the Nevada site an attractive test case for Bayesian framework proposed in section~\ref{sec: Bayesian Framework}. 
 \subsection{InSAR and GPS Data}
 \label{sec: GPS InSAR Data}

We generate an interferogram spanning May 4, 2003 and October 26, 2003 to measure the LOS deformation caused by the 2003 aquifer test. The LOS unit look vector is $[0.381, -0.08, 0.921]$. We observe a subsidence bowl southeast of the well WX31 and an uplift bowl northwest of the well (Figure~\ref{fig: Test Case}\emph{d}). This asymmetrical pattern suggests the possible existence of a fault as reported in \citeA{Burbey08}. Our interferogram shows similar deformation patterns as the one that spans May 4, 2003 and September 5, 2004 \cite{Burbey08}. This suggests that little deformation occurred after the end of the well test between October 26, 2003 and September 5, 2004, confirming that the regional GW recharge is minimal. 

We select a high phase-correlation ($>0.3$) block bounded by the gray outline in Figure~\ref{fig: Test Case}\emph{d} as the input for the inversion. In this 8-by-7 $\text{km}^{2}$ region, there are no visible orbital errors, ionospheric artifacts, or phase unwrapping errors that impact the deformation estimates. DEM errors and tropospheric errors are not substantial, because the study site is relatively flat with a dry desert climate. The dominant error source here is due to the change of surface scattering properties between the two SAR acquisitions. This phase noise (known as phase decorrelation) is not correlated in time or space \cite{zebker1992}. We estimate the phase decorrelation noise level ($\sim$ 3.2 mm) directly from the data and use it as $\sigma_\text{InSAR}$ in equation~\refp{equ: InSAR Misfit}.

We use the lateral displacements as recorded at 6 GPS stations during the first 22 days of the 2003 pumping test (Figure~\ref{fig: Test Case}\emph{e}) as an independent validation of the InSAR-based permeability estimates. These GPS stations are closer to the well, and little deformation was recorded at the other 4 stations that are further away. 
The joint inversion of GPS and InSAR data sets as well as their relative merits will be discussed in a future article.

\subsection{The Aquifer Model}
\label{sec: Aquifer Model}

We model the aquifer site as an 885 m deep layered poroelastic medium (as shown in Figure~\ref{fig: Test Case}\emph{c}) over the region bounded by the red polygon in Figure~\ref{fig: Test Case}\emph{a}. We assume no GW flux at all boundaries, zero displacement at the lateral boundaries, zero normal displacement at the bottom boundary, and a traction-free top surface. We set the initial deviation from the hydrostatic pressure before the start of the pumping, $p_0$, to zero. We model the groundwater extraction as a volumetric sink term, $f_p$ in equation~\refp{equ: PDE 3 Field a}, located in the segment of the well coinciding with the lower half of the aquifer (the blue segment of the well in Figure~\ref{fig: Test Case}\emph{c}).  Following \citeA{Burbey08}, the aquifer parameters are listed in Table~\ref{tab: Biot Model Parameters}. 

We set up the Mat\'ern class prior Gaussian field defined in equation~\refp{equ: Prior Elliptic SPDE}, which yields permeability samples with the following properties: (1) on average, at any given point in space, the permeability values vary from the prior mean listed in Table~\ref{tab: Biot Model Parameters}\emph{b} by a $\sim 1.5$ standard deviation (SD) in decimal logarithm; (2) the permeability samples correlate anisotropically in space, with a correlation length $\rho = 2$ km in the lateral direction and an order of magnitude larger correlation length in the vertical direction. The large vertical correlation length suppresses vertical variations that are not informed by the surface deformation data. Therefore, the estimated permeability should be interpreted as the vertical average of the permeability field. This average is a useful up-scaled representation of the aquifer permeability, because the GW flow is predominantly horizontal and along the unresolved fine-scale horizontal layering in the aquifer \cite{Cherry1979}.  In section~\ref{sec: Prior Role}, we show that the inverse solution is robust irrespective of a wide range of SD and $\rho$ in the horizontal direction.

\begin{table}[h]
\caption[The forward Model Parameters]{Forward model parameters} 
\begin{center}
\subcaption{Model parameters}
\begin{tabular}{ c|c }
\hline
 Volumetric force ($\mathbf{f_u}$)  &  0    \\
 Pumping rate ($-\int_\Omega f_p \mathrm{d}\mathbf{x}$)          &  $9{,}028$ $ \text{m}^3/ \text{day}$   \\
 Fluid viscosity ($\mu$)          &  $0.001$ $ \text{Pa}\cdot\text{s} $   \\
Biot-Willis coefficient ($\alpha$)       &  $0.998 $   \\
Height of the domain ($h$)            &  $885$ $\text{m} $  \\
Water density ($\rho$)         &  $997.97$ $\text{Kg}/\text{m}^3$ \\
Gravitational acceleration ($g$)            &  $ 9.8$ $\text{m}/\text{s}^2$ \\
Water compressibility  ($\beta_w$)      &  $ 4.4 \times10^{-10}$ $\text{Pa}^{-1}$ \\
 \hline
\end{tabular}\\[0.3in]
\subcaption{Layer parameters}
\begin{tabular}{ c|c|c|c|c }
\hline
                                             &  Units   & Aquifer                    & Confining layer        &  Lower clay          \\
\hline
Poisson's ratio ($\nu$)                      &    -            & $0.25$                     &  $0.25$                & $0.25$               \\
Drained shear modulus ($G$)                  &    Pa           & $3.4 \times 10^8$          & $3.5 \times 10^8$      & $8 \times 10^8$      \\ 
Specific storage ($S_\epsilon$)              & $\text{Pa}^{-1}$   & $2.1 \times 10^{-9}$       &  $1.2 \times 10^{-9}$  & $1.2 \times 10^{-9}$ \\ 
Prior mean value ($\text{log}_{10}(\kappa)$) &  $\text{m}^{2}$ (for $\kappa$)      &     -11.2                  &         -14.2          &     -14.2                  \\
\hline 
\end{tabular}
\end{center}
\label{tab: Biot Model Parameters}
\end{table}

\begin{figure}[h]
\centering
\includegraphics[width=1\textwidth]{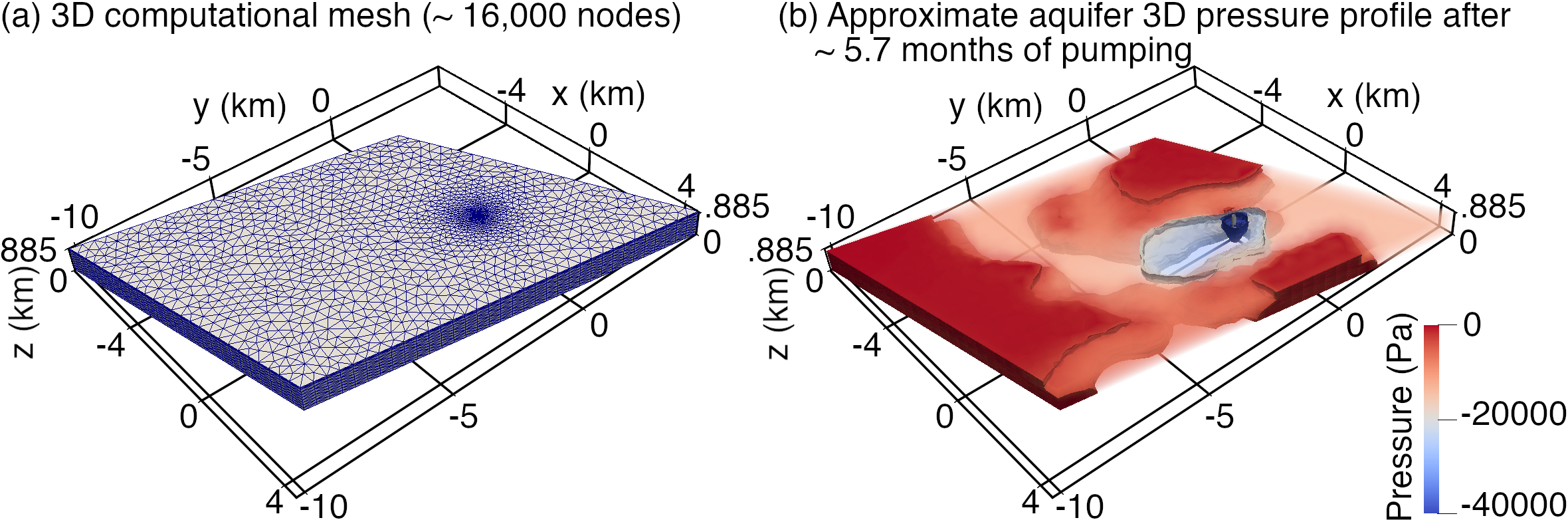}%
\caption{
        (a) The computational mesh of $16{,}896$ nodes. The mesh is generated using \texttt{Gmsh} and is refined around the well.
        (b) Simulated three-dimensional aquifer pore pressure profile at the time of the second InSAR data acquisition ($5.7$ months of pumping). This solution is obtained from the forward model using an InSAR-data-based MAP permeability estimate. The dark blue, gray, and red isovolumes correspond to the ranges $-40{,}000$ to $-39{,}900$, $-20{,}000$ to $-19{,}000$, and $-4{,}000$ to 0 (Pa), respectively. These isovolumes are cropped at $z=.8$ km and thus are shown only in the aquifer and the lower clay for better visualization.}
\label{fig: Pressure and Mesh}
\end{figure}

We generate two unstructured tetrahedral meshes of differing resolution for our computational domain using \texttt{Gmsh} \cite{GeuzaineRemacle09}. The number of nodes in these two meshes, and hence the number of DOFs approximating the log permeability $m$, are $4{,}081$ and $16{,}896$ (Figure~\ref{fig: Pressure and Mesh}\emph{a}). The overall number of DOFs at each time step of the discrete system state variables ($p$, $\mathbf{u}$ and $\mathbf{q}$) are $72{,}980$ and $320{,}824$, respectively. This system is built using the discretely-mass-conserving FEM described in section~\ref{sec: Forward Model}. The $4{,}081$-node mesh provides sufficient accuracy for our application and therefore we use this mesh in all results presented in this article except where noted. We construct the mesh so that the elements are refined near the well location and we ensure no two GPS station locations belong to the same element. To avoid excessive mesh refinement, we do not mesh to the exact well radius but rather to a cylindrical volume of radius 7 m that has the same axis as the well. The sink term is distributed over that volume such that it integrates to the pumping rate. We discretize the temporal evolution into 154 variable-length time steps over a period of 175 days. At the start of the simulation, the time step is set to 1.2 hours to accurately capture the rapid changes after the onset of pumping. The time step length increases gradually to up to 5-day time interval toward the end of the simulation.

We implement the forward model, adjoint system, and gradient and Hessian evaluations using the \texttt{Python}-based \texttt{FEniCS} library \cite{LoggMardalGarth12} for FEM discretization in space. We use the \texttt{hIPPYlib} library for state-of-the-art Bayesian and deterministic PDE-constrained inversion algorithms and prior construction \cite{VillaPetraGhattas2018}. Using this implementation we infer the log-permeability MAP point, i.e.\ the point in the parameter space that maximizes the posterior distribution, from the InSAR data. Based on this inferred permeability field, the forward model can simulate a three dimensional evolution of the pore pressure in response to the pumping test (e.g. Figure~\ref{fig: Pressure and Mesh}\emph{b}) and the expected deformation in east, north, and vertical directions at each time interval. We then compare this model-based lateral deformation with the GPS data set for validation of the inverse solution.

\section{Numerical Results}
\label{sec: Results} 
In section~\ref{sec: InSAR charcterization}, we present results of applying the Bayesian inversion framework to infer the permeability field of the Nevada test case described in section~\ref{sec: Test Case}. Section~\ref{sec: GN convergence} demonstrates the superiority of Newton-type methods over steepest descent for these poroelastic inverse problems.

\subsection{Characterizing the Aquifer Permeability Using InSAR Data}
\label{sec: InSAR charcterization}

\begin{figure}[htb]
    \centering
    \includegraphics[width=1\linewidth]{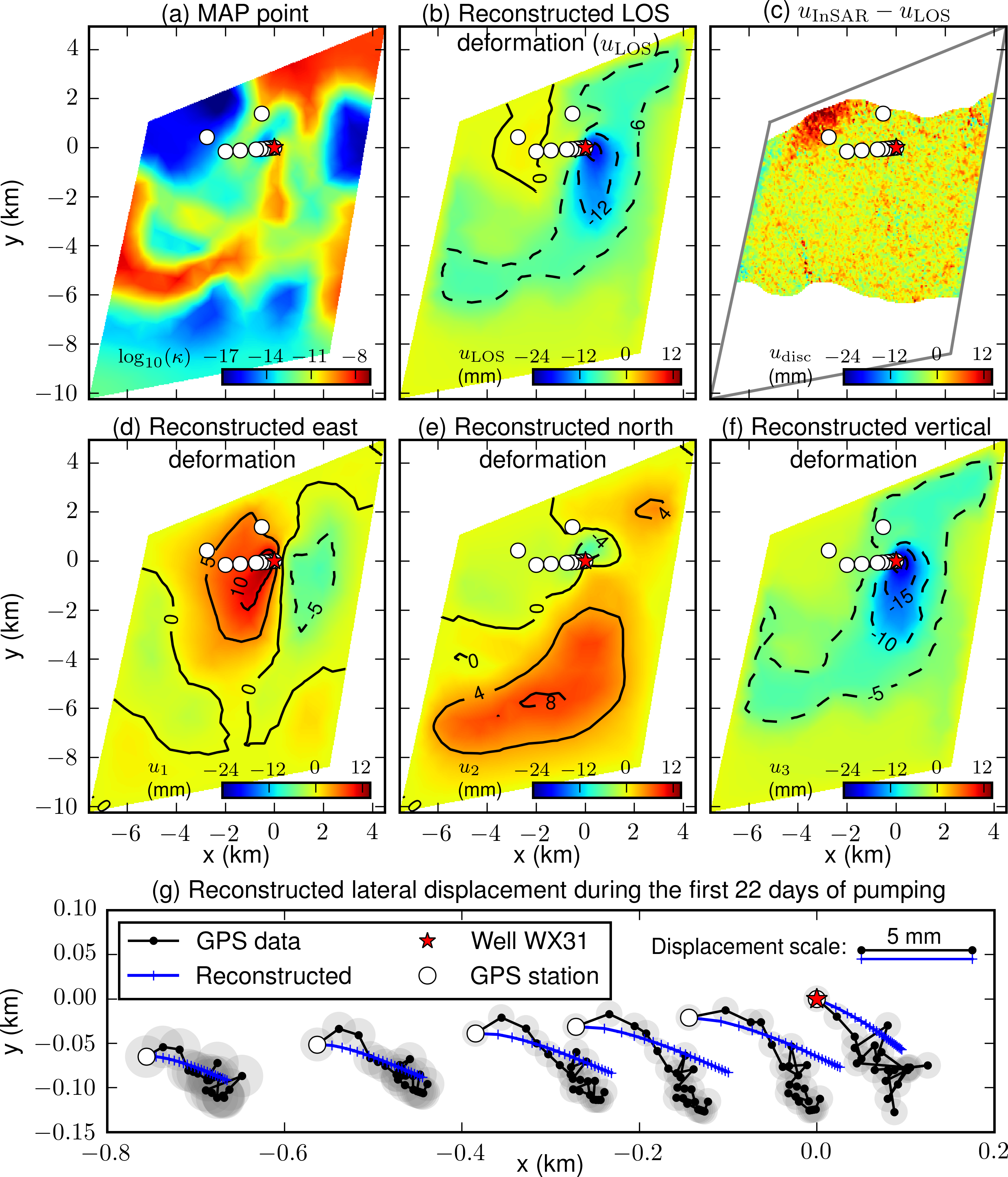}
    \caption{Results of characterizing Nevada aquifer permeability using InSAR data. In all the panels, the white circles mark the GPS station locations and the red star marks the well location. 
    (a) Two-dimensional horizontal slice of the inferred permeability, $e^{\mathbf{m}_\text{MAP}}$, at mid-aquifer depth (in decimal logarithm).
    (b) Reconstructed LOS displacement from the forward model using the permeability profile in Figure~\ref{fig: Aquifer Characterization}\emph{a}.
    (c) The discrepancy between InSAR data and the reconstructed LOS displacement: $u_\text{disc} = u_\text{InSAR}- u_\text{LOS}$.
    (d)--(f) Displacement components ($u_1$, $u_2$, and $u_3$) of the LOS deformation in Figure~\ref{fig: Aquifer Characterization}\emph{b}.
    (g) Map view of the reconstructed (blue) versus observed (black) cumulative lateral displacements at the GPS station locations from day $1$ to day $22$ of the pumping. At each GPS station, each black dot shows the daily cumulative lateral displacement. In all the panels, $x$ is the east direction and $y$ is the north direction relative to the well location (x,y)=(0,0)}
    \label{fig: Aquifer Characterization}
\end{figure}

We solve for the log-permeability MAP point, $\mathbf{m}_\text{MAP}$, using the Envisat interferogram that spans May 4, 2003 and October 26, 2003 (Figure~\ref{fig: Test Case}\emph{d}). The two-dimensional horizontal slice of the reconstructed permeability, $\kappa=e^{\mathbf{m}_\text{MAP}}$, at mid-aquifer depth (Figure~\ref{fig: Aquifer Characterization}\emph{a}) reveals distinct features in the permeability field: A high permeability channel extending from south/southwest of the well (marked by red star) to the west; and a low-permeability flow barrier northwest of the well.

Using this InSAR-inferred permeability field and the forward model, we simulate the expected LOS deformation that occurred between May 2003 to October 2003. This reconstructed LOS displacement accurately captures the subsidence bowl due to the pumping test. In most of the region, the discrepancy between the InSAR data and the simulated LOS deformation (Figure~\ref{fig: Aquifer Characterization}\emph{c}) is roughly random noise of magnitude within the estimated noise level. We observe a relatively large discrepancy in the northwest, likely due to the existence of a fault \cite{Burbey08} that is not accounted for in our aquifer model. In section~\ref{sec: Model Error}, we discuss the impact of the model error in our inversion solution.

Note that the simulated LOS deformation is computed from the simulated east, north, and vertical deformations (Figure~\ref{fig: Aquifer Characterization}\emph{d}--\ref{fig: Aquifer Characterization}\emph{f}) following equation~\refp{equ:los}. To validate our inversion results, we further compare the simulated lateral deformation with the lateral displacements as recorded at 6 GPS stations (Figure~\ref{fig: Aquifer Characterization}\emph{g}). We confirm that the reconstructed lateral displacements capture the GPS-observed nontrivial southeast-trending surface deformation during the first 22 days of the 2003 pumping test. It is worth noting that the northward component of these GPS-observed lateral displacements is well-approximated by our model despite the fact that the northward contribution to the LOS deformation is negligible ($\alpha_2 = -0.08$). This can be achieved only when a high fidelity physical approximation---the 3D linear poroelasticity assumption---is incorporated into the inversion framework.

\subsection{Convergence of the (Gauss--)Newton Method}
\label{sec: GN convergence}

\begin{figure}[htb]
    \centering
    \includegraphics[width= 1\linewidth]{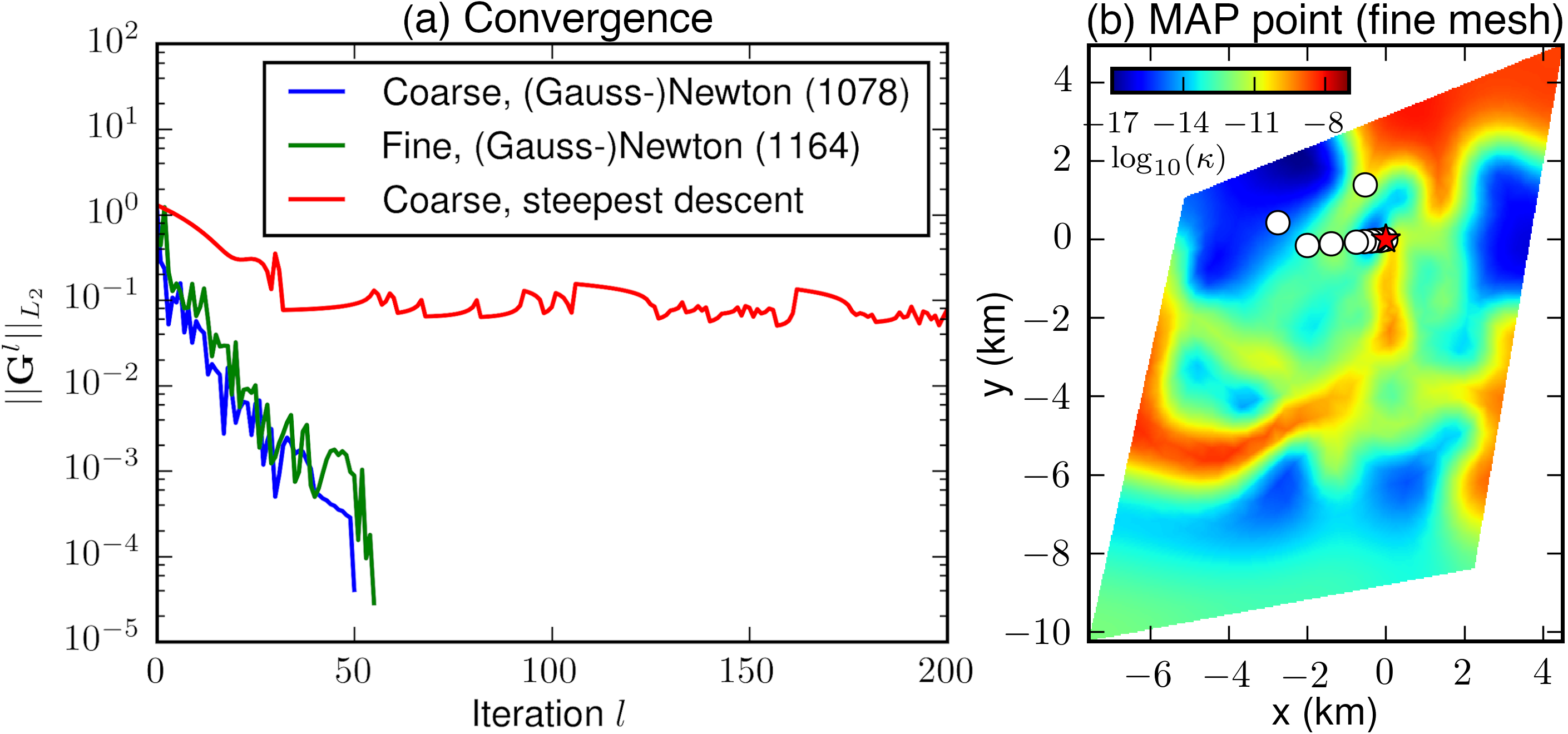}
    \caption{(a) Convergence of the (Gauss--)Newton method for finding the InSAR-data-based MAP point using the coarse mesh, $4{,}081$ nodes, (blue) and the fine mesh, $16{,}896$ nodes, (green). The total number of CG iterations for each case is provided in parentheses in the legend.  The convergence criterion we set is reducing the L\textsuperscript{2} norm of the gradient by four orders of magnitude.
    (b) A two-dimensional horizontal slice of the InSAR-data-based MAP point, $e^{\mathbf{m}_\text{MAP}}$, at mid-aquifer depth (in decimal logarithm) using the $16{,}896$ node mesh. }
    \label{fig: Convergence}
\end{figure}

The (Gauss--)Newton conjugate gradient method, described in section~\ref{sec: MAP}, is known to converge rapidly when applied to optimization problems for which the data misfit Hessian effective rank is relatively small and independent of the discretization dimension \cite{Heinkenschloss93}. As has been mentioned in section \ref{sec: MAP}, this has been demonstrated either analytically or numerically for a wide spectrum of geophysical inverse problems. Although this property of the Hessian has not been shown analytically for poroelastic inverse problems, our numerical results confirm previous work by \citeA{HesseStadler14} that demonstrated the computational advantage of Newton-type methods, suggesting rapid decay of the eigenvalues of the data misfit Hessian for this problem.

To verify the dimension-independent convergence of (Gauss--)Newton--CG iteration, in the form of equation~\refp{equ: Newton Iteration}, we solve for the permeability MAP point for both our coarse and fine meshes and compare their convergence in the blue and green curves in Figure~\ref{fig: Convergence}\emph{a}). In both cases, the (Gauss--)Newton--CG method converges in a small number of iterations (less than 60). Even though the fine mesh has four times the number of parameter DOFs as the coarse mesh, the number of iterations and PDE solves required for both meshes are comparable (Table~\ref{tab: Convergence}). This demonstrates that the algorithm is dimension independent and hence scalable to larger problems. This is essential if the InSAR-based Bayesian poroelastic aquifer characterization is to be applied to GW management at the basin scale \cite{Chen2016,Chen2017}. Additionally, the inferred permeability fields in both the coarse mesh (Figure~\ref{fig: Aquifer Characterization}\emph{a}) and the fine mesh (Figure~\ref{fig: Convergence}\emph{b}) are consistent. Note that the method was set to use the Gauss--Newton Hessian in the first 50 iterations, then switch to using the full Newton Hessian  in the remaining iterations.

We also demonstrate the superiority of the (Gauss--)Newton--CG method's convergence over the classical steepest descent method on the coarse mesh (red line in Figure~\ref{fig: Convergence}\emph{a}). As shown in Table~\ref{tab: Convergence}, the steepest descent method terminated after 18 days of runtime (due to hitting the iteration limit of $5{,}000$) without converging. In contrast, the (Gauss--)Newton--CG method converged in $4.5$ hours, giving an apparent speedup of $100\times$ (but the true speed up would be much larger had we allowed steepest descent to run to convergence). 
Comparing the number of PDE solves in each case, we see that steepest descent method required 30 times the number of PDE solves required by the (Gauss--)Newton--CG method and yet did not converge (Table~\ref{tab: Convergence}). 

We point out that part of the computational savings obtained when using the {(Gauss--)}Newton--CG method are due to the cost of a PDE solve for the steepest descent method being greater than that for the (Gauss--)Newton--CG method in our implementation. On average, a PDE solve requires 20 seconds in the former case, while requiring only 7 seconds for the latter case on the coarse mesh. This is a result of using a direct solver for the linear poroelasticity equations. When using the GN method, the direct solver's LU factors can be computed once at the $l^\text{th}$ GN iteration, stored, and reused in all the CG iterations required at that GN iteration. This is because the same incremental forward and adjoint PDEs are solved at each CG iteration (equations~\ref{equ: Incremental Forward} and \ref{equ: Incremental Adjoint}). On the other hand, for the steepest descent method, the parameter $m^l$ changes in each iteration and thus a new factorization of the poroelasticity operator is required at each steepest descent iteration.

\FloatBarrier

\begin{table}[h]
    \caption{Convergence results for computing the MAP point for three cases. The first two rows report convergence results of the (Gauss--)Newton method using the coarse ($4{,}081$ nodes) mesh and the fine ($16{,}896$ nodes) mesh, respectively. The third row reports convergence results of steepest descent method (SDM) using the coarse mesh. The gradient reduction factor is defined as $\text{initial gradient norm}/ \text{final gradient norm}$.}
    \begin{center}
        \begin{tabular}{c|c|c|c|c|c}
            \hline
              Method,    & Gradient        & PDE solves$^a$:       & Num. of global & Num. of total  & Total \\
              mesh      & reduction       &  forward (fwd),  & iterations    & CG it. & time \\
                        &factor             & adjoint (adj)  &        &        &  \\
            \hline
            GN,         & $\geq 10^4$     & fwd: $107 + 1,078$,     &    50        & $1,078$   & 4.5h  \\
            coarse      &                 & adj: $51 + 1,078$     &              &        & \\
            \hline
            GN,         & $\geq 10^4$     &fwd: $103+ 1,164$,     &    55        & $1,164$   & 22.3h \\
            fine        &                 &adj: $56 +1,164$      &              &        & \\
            \hline
            SDM,         & $\approx 6.4\text{e}2$ &fwd: $66,000$,     & $5,000$         & -      & 18d   \\
            coarse      &                 & adj: $5,000$          &              &        & \\
            \hline
            \multicolumn{6}{p{13cm}} {$^{a}$  The number of forward and adjoint PDE solves are given as: number of forward (or adjoint) poroelasticity solves $+$ number of incremental forward (or adjoint) poroelasticity solves.}
        \end{tabular}
    \end{center}
    \label{tab: Convergence}

\end{table}

\section{Discussion}
\label{sec: Discussion}
In this section, we discuss how the essentially arbitrary pixel spacing of the InSAR data can be treated systematically in the Bayesian setting, without the introduction of artificial weights. We then show that the inferred MAP point of the permeability field is not appreciably affected by the choice of prior. We further show that the discrepancy between the model and the data beyond the noise identifies a region where our model exhibits structural error. We conclude this section with a brief overview of Part II of this study, which focuses on MCMC sampling of the Bayesian posterior, building on tools developed in the present article.

\subsection{Multi-looking InSAR Data with Adjusted Noise Estimation}
\label{sec: InSAR Resolution}

\begin{figure}[h]
    \centering
    \includegraphics[width=1\linewidth]{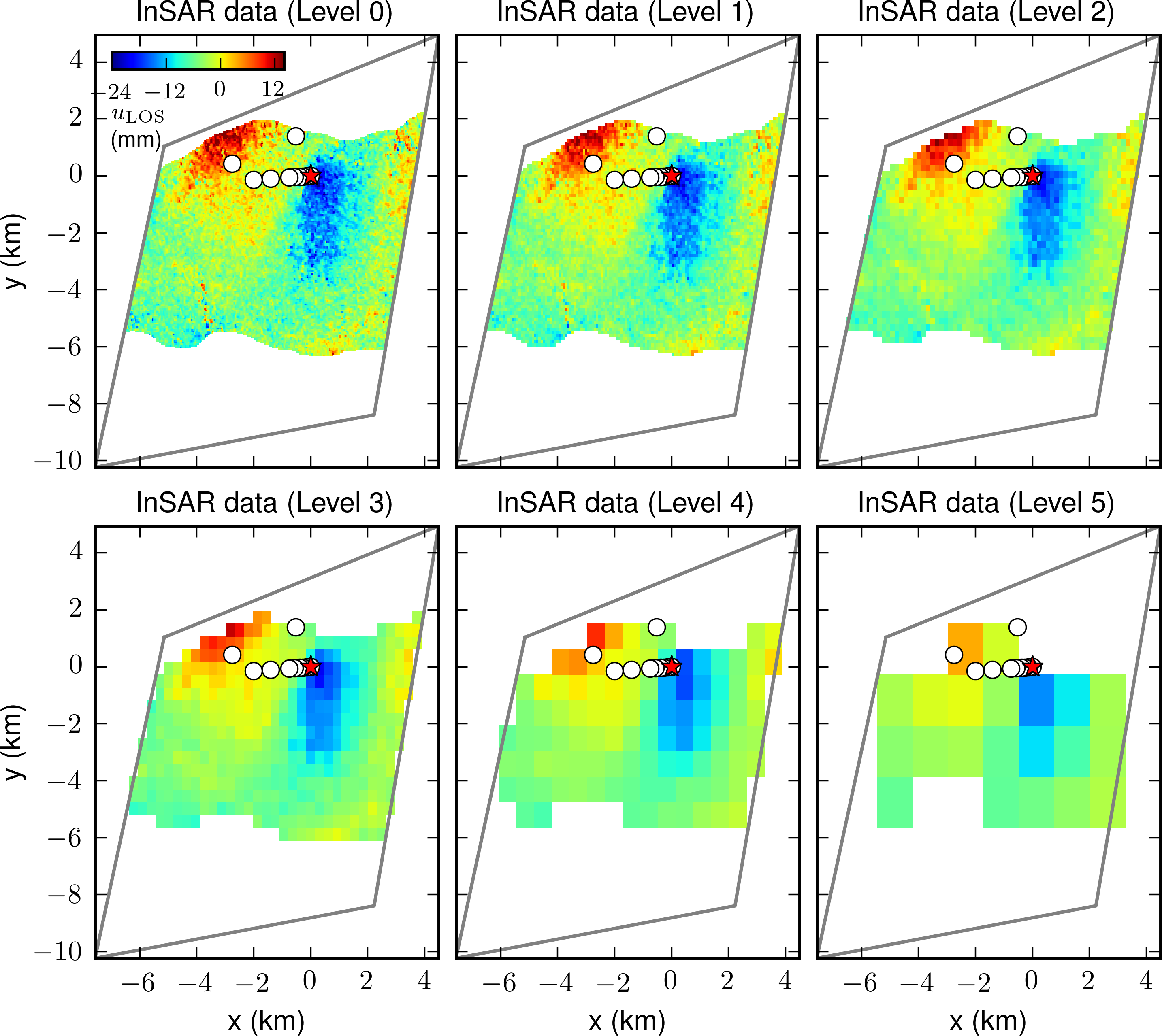}
    \caption{Downsampled InSAR data sets labeled by the downsampling level. At each downsampling level, adjacent two-by-two data points are averaged into a single data point. Level zero is the original data set ($31{,}003$ data points). The number of data points are $7{,}679$, $1{,}891$, 452, 101, and 21 data points for the levels 1, 2, 3, 4, and 5, respectively.}
    \label{fig: Downsampled InSAR Data}
\end{figure}
  
\begin{figure}[h]
    \centering
    \includegraphics[width=1\linewidth]{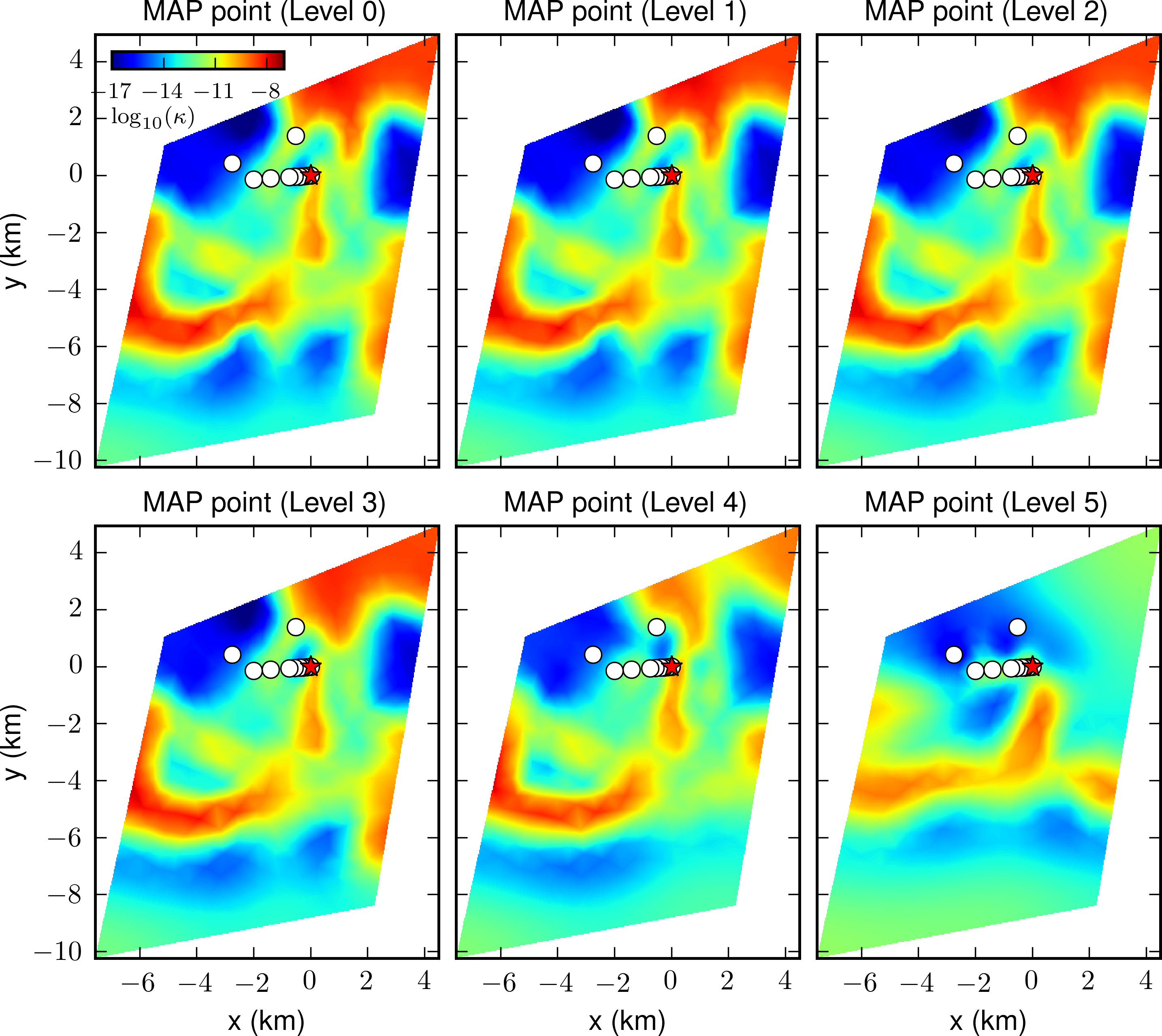}
    \caption{Two-dimensional horizontal slices of the permeability MAP point, $e^{\mathbf{m}_\text{MAP}}$, at mid-aquifer depth (in decimal logarithm) inferred from each of the six downsampled InSAR data sets in Figure~\ref{fig: Downsampled InSAR Data}.}
    \label{fig: MAP from Downsampled Data}
\end{figure}

Here we study the sensitivity of the inverse solution to the essentially arbitrary pixel spacing of the InSAR data. We downsample the original InSAR data multiple times (Figure~\ref{fig: Downsampled InSAR Data}). Each level of downsampling is performed by averaging adjacent two-by-two pixels into a single pixel. This is known as multi-looking, which reduces the InSAR phase decorrelation noise by a factor of two. We incorporate this noise adjustment in \eqref{equ: InSAR Misfit}. Figure~\ref{fig: MAP from Downsampled Data} shows the resulting inferred permeability MAP point at mid-aquifer depth using each of the six multi-looking InSAR data. When the number of InSAR pixels is sufficient to resolve the main deformation features (Level 0-3), the inferred permeability solutions are similar. Hence we conclude that the InSAR data redundancy can be addressed in a fully Bayesian setting, i.e., without introducing artificial weights, by treating the data noise appropriately. However, if InSAR data are too coarse to resolve the main deformation features (Level 4-5), a deterioration in the validity of the inferred permeability field is expected.

\subsection{The Role of the Prior}
\label{sec: Prior Role}

The limited data available for aquifer characterization has placed emphasis on methods to inform the characterization with prior geostatistical knowledge \cite{Goovaerts1997,Deutsch1998,Wackernagel2010}. A large number of methods that allow imposition of additional knowledge from small-scale horizontal layering to the connectivity of large-scale fluvial channels \cite{Remy2009,Mariethoz2014} have been developed. Unfortunately, most of these methods are not compatible with the scalable framework used here that requires a sparse prior precision operator. However, we show below that InSAR data are sufficiently informative that dense geostatistical priors are not required. 

\begin{figure}[h]
    \centering
    \includegraphics[width=1\textwidth]{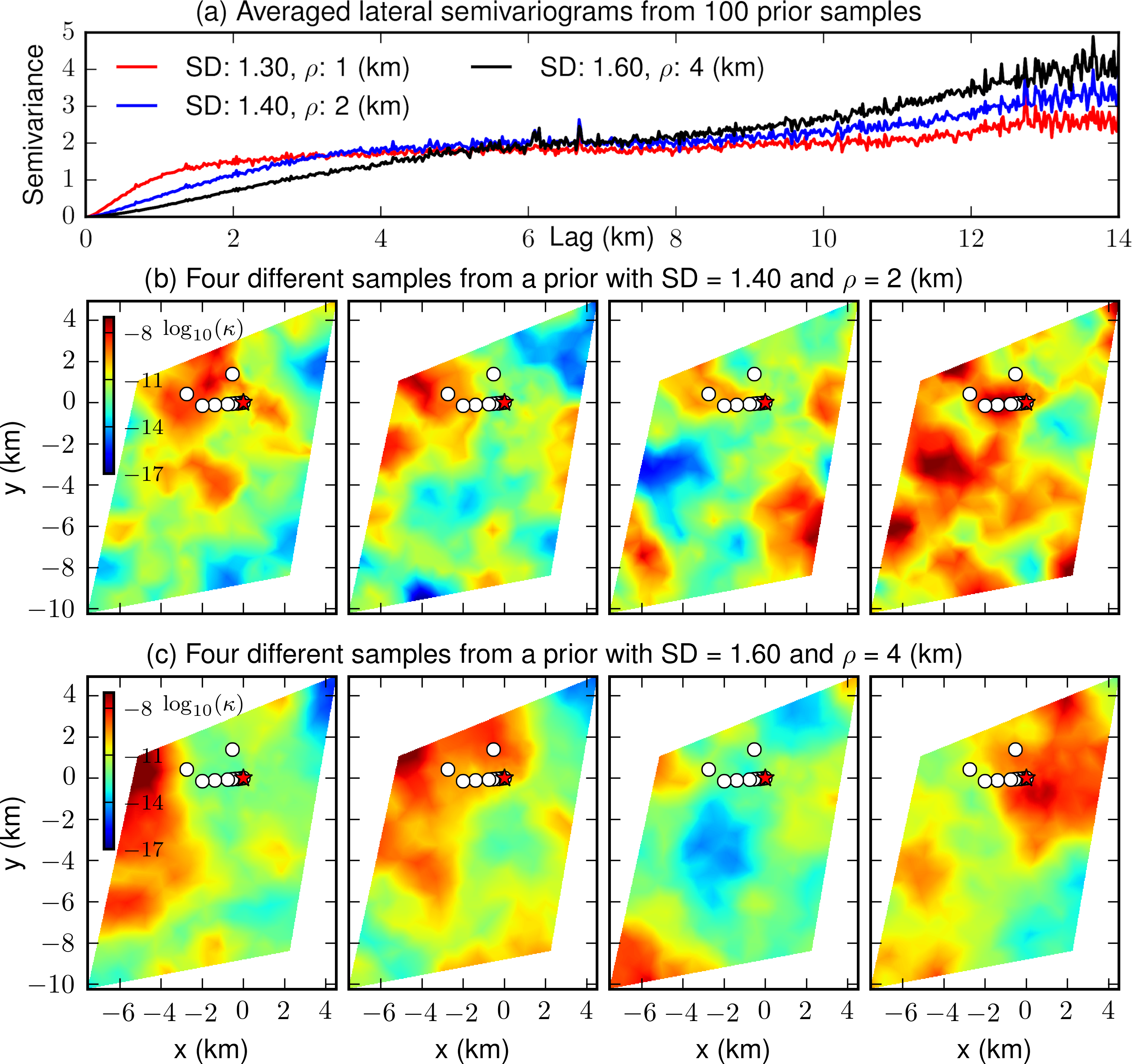}%
    \caption{(a) Averaged semivariance over 100 samples from the three prior choices with $\rho = 1$, $2$, and $4$ and an average-over-domain SD $\approx 1.5$ (reported in Table~\ref{tab: Prior}). (b, c) Two-dimensional horizontal slices of prior samples at mid-aquifer depth for two priors with different $\rho$ and similar SD. 
    }
    \label{fig: Prior Samples}
\end{figure}

Here we limit ourselves to the class of Mat\'ern covariances discussed in section~\ref{sec: Prior Matern}. These priors have the typical parameters of a semi-variogram \cite{Deutsch1998}, the range, $\rho \propto  \sqrt{\delta/\gamma}$, and the sill (Figure~\ref{fig: Prior Samples}\emph{a}). The sill is the square of the pointwise marginal standard deviation $\mathrm{SD}\propto\sqrt{1/(\delta\gamma)}$. Due to the boundary effects, the exact sill cannot be determined a~priori and is hence difficult to control exactly in Mat\'ern priors. Due to the continuous nature of the prior there is no nugget. In addition, the Mat\'ern prior also allows us to specify a mean, which is taken from \citeA{Burbey08}. Two-dimensional horizontal slices of prior samples with SD$\approx 1.5$ and increasing range from $\rho=2$ to $\rho=4$, are shown in Figure~\ref{fig: Prior Samples}\emph{b} and \ref{fig: Prior Samples}\emph{c}, respectively. We also note that in infinite dimensional inverse problems the prior must act as a regularization.  Table~\ref{tab: Prior} notes priors that do not regularize the inverse problem sufficiently so that MAP point estimate does not converge to the required tolerance.

\begin{table}[h]
    \caption{ Parameters for five different prior choices. The SD values are reported as three values: minimum (third column), maximum (fourth column), and average over domain (fifth column). The values of $\gamma_1$ and $\delta_1$ are $3.33$ and $3.33\mathrm{e}{-6}$, respectively. The second to last column reports whether inverting for the MAP point has converged (yes) or stopped when reaching the maximum number of backtracking steps (no). The convergence criterion we impose is the reduction of the L\textsuperscript{2} norm of the gradient by 4 orders of magnitude.} 
    \begin{center}
        \begin{tabular}{r|r|l|c|c|c|c||c|c}
            \hline 
            $\gamma$                & $\delta$               & SD   &SD   & SD  & $\rho$ & $\sqrt{1/(\delta\gamma)}$     & Converged & Gradient\\
                                    &                        & min.\  &max.\  & avg.\ & (km)   &                         &           & reduction\\
            \hline
            $\gamma_1$              & $\delta_1$             & 1.1  & 3.0 &1.4  & 2      & 300                           & yes       & $\geq 10^4$\\
            \hline
            $\frac{1}{2}\gamma_1$   & $\frac{1}{2}\delta_1$  & 2.3  & 6.1 &2.8  & 2      & 600                           & no        & $\approx 1.8\mathrm{e}3$\\
            $2\gamma_1$             & $2\delta_1$            & 0.59 & 1.5 &0.7  & 2      & 150                           & yes       & $\geq {10}^4$\\
            \hline   
            $\frac{1}{2}\gamma_1$   & $2\delta_1$            & 1.1  & 3.0 &1.3  & 1      & 300                           & no        & $\approx 3.1\mathrm{e}3$\\
            $2\gamma_1$             & $\frac{1}{2}\delta_1$  & 1.2  & 3.0 &1.6  & 4      & 300                           & yes       & $\geq 10^4$\\
            \hline
        \end{tabular}
    \end{center}
    \label{tab: Prior}
\end{table}
 
\begin{figure}[h]
    \includegraphics[width=1\linewidth]{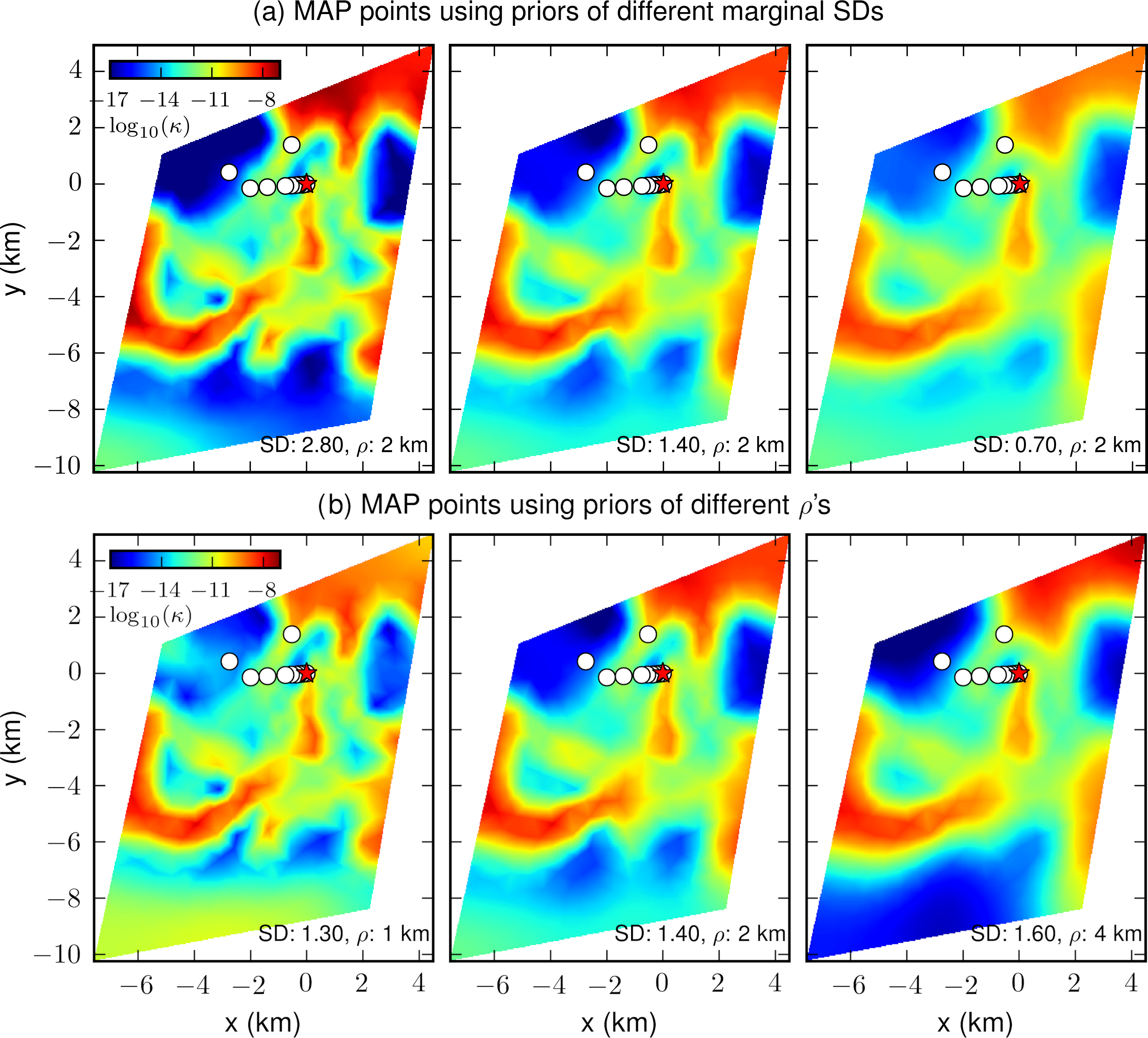}
    \caption{Two-dimensional horizontal slices of the permeability MAP point, $e^{\mathbf{m}_\text{MAP}}$, at mid-aquifer depth (in decimal logarithm) for the five chosen priors in Table~\ref{tab: Prior}. To make the visual comparison easier, the MAP point for which (avg.\ SD, $\rho$) = (1.4, 2 km) is shown twice: in the middle of the top and bottom rows.
    (a) The average-over-domain SD decreases from left to right: $2.8$, $1.4$,  to  $0.7$, while $\rho$ = 2 is constant.
    (b) $\rho$ increases from left to right: $1$, $2$, to $4$ kilometers, while the average-over-domain SD $\approx1.5$ is almost constant.}
    \label{fig: MAP Using Different Prior}
\end{figure}

The high spatial resolution of the InSAR data provides good constraints on the lateral permeability variation in the aquifer. This can be seen in Figure~\ref{fig: MAP Using Different Prior}, which shows the MAP points for each of the five prior choices in Table~\ref{tab: Prior}. Despite large ranges in both $\rho$ and SD, the main features of the MAP point remain almost unchanged. This includes both the high permeability channel south of the well and the low-permeability region northwest of the well. Of course, the amplitudes increase with SD and small features are lost as $\rho$ increases, but main pattern is independent of the prior. This demonstrates that the MAP point is dominantly informed by the data and that the effect of the prior on the inferred permeability variation is small. 

This highlights the step change in the data availability for the characterization of lateral aquifer heterogeneity that is provided by InSAR. Given upcoming missions with improved accuracy and increased frequency, InSAR-derived surface deformation measurements will provide a powerful and low-cost means of aquifer characterization.

\subsection{Assessment of the Model Error}
\label{sec: Model Error}

\begin{figure}[h]
    \includegraphics[width=1\linewidth]{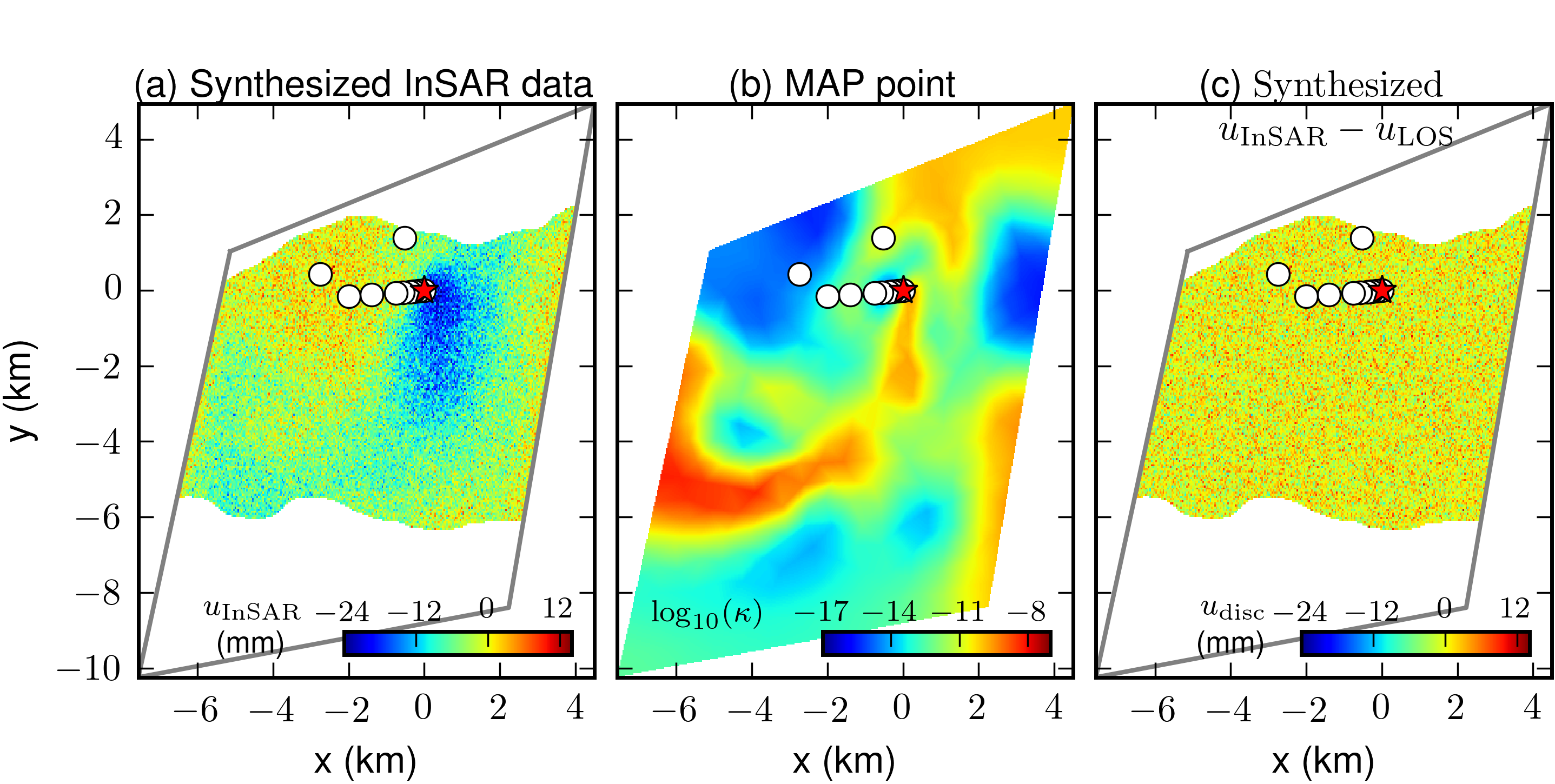}
    \caption{(a) Synthesized InSAR data generated from the forward model with the permeability field in Figure~\ref{fig: Aquifer Characterization}\emph{a}.   (b) Two-dimensional horizontal slice at mid-aquifer depth of the permeability, $e^{\mathbf{m}_\text{MAP}}$, inferred using the data in Figure~\ref{fig: Synthesized Aquifer Characterization}\emph{a} (in decimal logarithm). (c) The discrepancy between the synthesized InSAR data and the reconstructed LOS displacement: $u_\text{disc} = u_\text{InSAR}- u_\text{LOS}$, where $u_\text{LOS}$ is reconstructed from the model with the permeability field in Figure~\ref{fig: Synthesized Aquifer Characterization}\emph{b}.}
    \label{fig: Synthesized Aquifer Characterization}
\end{figure}

In section~\ref{sec: InSAR charcterization}, we showed that the discrepancy between the model and the data is roughly a random noise signal of magnitude within the estimated noise level, except in the northwest corner of the domain
(Figure~\ref{fig: Aquifer Characterization}\emph{c}). \citeA{Burbey08} have postulated the existence of a fault in this area, which is not accounted for in our aquifer model, presented in section~\ref{sec: Aquifer Model}. Here we investigate whether the relatively large discrepancy in the northwest corner is due to this structural error in the model.

Introducing the actual fault and the associated mechanics is beyond the scope of this article. Instead, we explore the possibility of model error by studying a synthetic inversion. For this model-error-free scenario we regard the permeability field inferred from the real InSAR data (Figure~\ref{fig: Aquifer Characterization}\emph{a}) as the ``true'' permeability. Using this truth permeability field, a synthesized InSAR LOS displacement map is generated from the model-predicted LOS observations over the same period and area as for the actual InSAR observations. These synthesized observations are then polluted by a normally-distributed random noise of the same magnitude assumed for the real InSAR data, $3.2$ mm (the synthesized data are shown in Figure~\ref{fig: Synthesized Aquifer Characterization}\emph{a}).

We then use the synthesized LOS data to solve the inverse problem of estimating the permeability MAP point.  The overall pattern of the resulting permeability MAP point (Figure~\ref{fig: Synthesized Aquifer Characterization}\emph{b}) is very similar to those of the ``true'' medium (Figure~\ref{fig: Aquifer Characterization}\emph{a}). There is, however, a notable decrease in the magnitude of the permeability variation from the prior mean,  due to the regularizing action of the prior. 

Using the permeability field inferred from the synthetic data (Figure~\ref{fig: Synthesized Aquifer Characterization}\emph{b}), we solve the forward problem and reconstruct the LOS observations, $u_\text{LOS}$. The error in these reconstructed LOS observations  (Figure~\ref{fig: Synthesized Aquifer Characterization}\emph{c}) is within the noise everywhere in the observed area. This demonstrates that we should be able to reconstruct the LOS observations to within data noise in the absence of model error. This suggests that the inability of the poroelatic model to reconstruct the LOS observations in the northwest (Figure~\ref{fig: Aquifer Characterization}c) is likely due to a number of possible simplifications in the model, including the presence of a fault.

Again this demonstrates the high information content of the InSAR data. Not only do they provide detailed constraints on the lateral permeability variation, they also clearly highlight regions where the model likely has structural errors. This allows the targeted improvement of the aquifer model.

\subsection{Exploration of the posterior distribution (Part II)}
\label{sec: Future Directions}
Our focus in part I of this two-part series of articles is on estimating the permeability maximum A posteriori point. In the preceding discussion, we have paid careful attention to matters arising in the Bayesian formulation of this inverse problem, including treatment of multilooking data sets, the construction of the prior and its influence on the MAP point, and analysis of model error. Our ultimate goal is to explore the posterior distribution and estimate the expected value along with the associated uncertainties of various quantities of interest. Estimating the MAP point is a critical first step because of its intrinsic importance as the most likely permeability realization. Furthermore, the MAP point can be used to create proposals for MCMC sampling algorithms required to explore the posterior distribution \cite{PetraMartinStadlerEtAl14, MartinWilcoxBursteddeEtAl12}. Since this discussion of estimating the MAP point is self-contained and draws important scientific conclusions about applying our method to the Nevada test case and the informativeness of InSAR data, and to avoid an unnecessarily lengthy presentation, we defer the discussion of the uncertainty quantification and MCMC sampling to part II.

\FloatBarrier
\section{Conclusion}
\label{sec: Conclusion}
We have shown that InSAR-based surface deformation measurements provide detailed information about lateral variations in aquifer permeability. In the Nevada test case, the surface deformation due to a single well allowed us to identify both high permeability pathways and flow barriers. Due to the high spatial resolution of the InSAR data, the main features of the inferred MAP permeability field are essentially independent of prior assumptions. The inferred permeability field also allows us to reconstruct the lateral deformations recorded at six GPS stations near the well. This provides validation, because the GPS data were not used in the inference of the permeability field. 

The above conclusions are derived from a single interferogram that captures the total subsidence due to the well test. Given the increasing availability of InSAR data from Sentinel-1 and upcoming NISAR missions, future aquifer characterization will be able to utilize higher quality and more frequent surface deformation measurements. We therefore believe that  geodetic surface deformation measurements will dramatically increase the information for aquifer characterization. Scalable and robust inversion frameworks that integrate InSAR data into poromechanical aquifer models are likely to become critical tools for regional groundwater management.

The Bayesian inference framework was applied to a three-dimensional transient multi-physics problem with real data that requires the inference of up to $16{,}896$ parameters. This requires dimension invariant methods and was achieved by exploiting the compact nature of the parameter-to-observable map, adjoint-based derivatives, and Gauss--Newton--CG method, the combination of which capture the intrinsic low dimensionality of this inverse problem and permit rapid, dimension-independent convergence. The power of these methods is available through the \texttt{Python}-based Bayesian inverse problems library \texttt{hIPPYlib}. The scalability of this framework provides the basis for Bayesian uncertainty quantification, based on Markov chain Monte Carlo sampling, which we will explore in Part II of this two-part series of articles.

\appendix
\section{Variational Formulation and Discretization Details of the 3-Field Formulation of the Biot System }
\label{app: MFEM Discretization}
For the variational form of the system~\refp{equ: PDE 3 Field a}--\refp{equ: PDE 3 Field c} to be well-defined, we assume the state variables $p(.;t)$, $u(.;t)$ and $\mathbf{q}(.;t)$ belong to the following infinite dimensional function spaces (note that these are regularity assumptions in space, we additionally require sufficient regularity in time):
\begin{align}
\mathcal{P} = &\big\{  p : \Omega \rightarrow {\rm I\!R} \quad | \; p \in L^2(\Omega)  \big\} \\
\mathcal{U} = &\big\{  \mathbf{u} : \Omega \rightarrow {\rm I\!R}^d \;\, | \; \mathbf{u} \in \left( H^1(\Omega) \right)^d 
                      \; \text{and} \; \mathbf{u}(x;t) =\mathbf{u_d} \; \text{on} \; \partial\Omega^d_u   \big\} \\
\mathcal{Q} = &\big\{  \mathbf{q} : \Omega \rightarrow {\rm I\!R}^d \;\, | \; \mathbf{q} \in  H(\text{div};\Omega) 
                      \; \text{and} \; \mathbf{q}(x;t)\cdot n =g \; \text{on} \; \partial\Omega^n_p \big\} ,
\end{align}
respectively, where $d =1$, $2$, or $3$ is the dimension of the physical domain. The space $ H(\text{div};\Omega)$ is defined as $  H(\text{div};\Omega)= \left\{f: f \in  \left( L^2(\Omega)\right)^d \; \text{and} \; \nabla \cdot f \in  L^2(\Omega)\right\} $. 

To obtain the weak form of the system~\refp{equ: PDE 3 Field a}--\refp{equ: PDE 3 Field c}, we multiply the equations~\refp{equ: PDE 3 Field a}, \refp{equ: PDE 3 Field b} and \refp{equ: PDE 3 Field c} by test functions $r \in \mathcal{P}_0$, $\mathbf{v} \in \mathcal{U}_0$ and $\mathbf{w} \in \mathcal{Q}_0$, respectively, where the function spaces $\mathcal{P}_0$, $\mathcal{U}_0$, and $\mathcal{Q}_0$ are defined as follows:
\begin{align}
\mathcal{P}_0 =& \big\{  p : \Omega \rightarrow {\rm I\!R} \quad | \; p \in L^2(\Omega)  \big\} \\
\mathcal{U}_0 =& \big\{  \mathbf{u} : \Omega \rightarrow {\rm I\!R}^d \;\, | \;
 \mathbf{u} \in \left( H^1(\Omega) \right)^d \; \text{and} \; \mathbf{u}(x;t) 
=\mathbf{0} \; \text{on} \; \partial\Omega^d_u   \big\} \\
\mathcal{Q}_0 =& \big\{  \mathbf
{q} : \Omega \rightarrow {\rm I\!R}^d \;\, | \; \mathbf{q} \in  H(\text{div};\Omega)  
  \; \text{and} \; \mathbf{q}(x;t)\cdot n =0 \; \text{on} \; \partial\Omega^n_p \big\} .
\end{align}
We then integrate the three equations in space over the domain $\Omega$. We discretize in time using backward Euler method. The variational problem is therefore the problem of finding $p_k \in \mathcal{P}$, $\mathbf{u}_k \in \mathcal{U}$ and $\mathbf{q}_k \in \mathcal{Q}$ that satisfy the following equations:
 \begin{align}
&\int_\Omega \big( \left(S_\epsilon p_k + \alpha \nabla \cdot \mathbf{u}_k \right) r + {\Delta t}_k \, \nabla \cdot \mathbf{q} \, r \big) \,\mathrm{d}\mathbf{x} = 
\int_\Omega ( {\Delta t}_k \, f_p + S_\epsilon p_{k-1} + \alpha \nabla \cdot \mathbf{u}_{k-1} )r\,\mathrm{d}\mathbf{x} \label{equ: Weak Form 3 Field a}\\
&-\int_\Omega \left(\pmb{\sigma}(\mathbf{u}_k) - \alpha p_k \mathbf{I} \right):\nabla \mathbf{v} \,\mathrm{d}\mathbf{x} = -\int_\Omega \mathbf{f_u} \cdot \mathbf{v} \,\mathrm{d}\mathbf{x}
  - \int_{{\partial\Omega}^n_\mathbf{u}} \mathbf{g}\cdot \mathbf{v} \,\mathrm{d}s \label{equ: Weak Form 3 Field b}  \\
&-\int_\Omega {\Delta t}_k \, (\frac{e^m}{\mu})^{-1} \mathbf{q}_k \cdot \mathbf{w} \,\mathrm{d}\mathbf{x} + \int_\Omega {\Delta t}_k \, p_k \nabla \cdot \mathbf{w} \,\mathrm{d}\mathbf{x}  = \int_{{\partial\Omega}^d_p} {\Delta t}_k  \, p_d \mathbf{w}\cdot n \,\mathrm{d}s ,  \label{equ: Weak Form 3 Field c}
\end{align}
for all the choices of test functions $r \in \mathcal{P}_0$, $\mathbf{v} \in \mathcal{U}_0$ and $\mathbf{w} \in \mathcal{Q}_0$ and for time steps $k =1,2, 3, .., N_t$. The functions $p_0$ and $\mathbf{u}_0$  are the initial values for the pressure and the displacement, respectively. Requiring that  $f_p(.,t) \in L^2(\Omega)$, $p_d(.,t) \in L^2({\partial\Omega}^d_p)$, $\mathbf{f}_\mathbf{u}(.,t) \in (H^{-1}(\Omega))^d$, and $\mathbf{g}(.,t) \in (H^{-1}(  {\partial\Omega}^n_\mathbf{u}))^d$ is necessary for the well-posedness of the weak problem. We follow \cite{FerronatoCastellettoGambolati10} in our choice of the finite element function spaces in which the pressure $p$ is approximated by piecewise constant function $\hat{p}(x,t)$ in $\mathcal{\hat{P}} \subset \mathcal{P}$ and the velocity is approximated in the lowest order Raviart-Thomas space by the function $\hat{q}(x,t)$ in $\mathcal{\hat{Q}} \subset \mathcal{Q}$. The displacement is approximated by first-order Lagrange polynomials, function  $\hat{\mathbf{u}}(\mathbf{x},t)$ in $\mathcal{\hat{U}} \subset \mathcal{U}$. This choice of elements results in continuity of normal velocity across the elements facets and hence guarantees discrete mass conservation. 

The discretized state functions can be written as follows (we adopt some of the notation from \cite{FerronatoCastellettoGambolati10}):
\begin{linenomath}
 \[\arraycolsep=5.0pt\def\arraystretch{1.0} \mathbf{u}(\mathbf{x},t) \approx \hat{\mathbf{u}}(\mathbf{x},t)  =  \left( \begin{array}{c}
\hat{u}_x(\mathbf{x},t) \\
 \hat{u}_y(\mathbf{x},t) \\
 \hat{u}_z(\mathbf{x},t) 
 \end{array} \right)  \]
\end{linenomath}
\begin{linenomath}
\[\arraycolsep=5.0pt\def\arraystretch{1.0}   = \left( \begin{array}{c}
\sum_{i=1}^{n_n}\phi_i(\mathbf{x}) u_{x,i}(t) \\
\sum_{i=1}^{n_n}\phi_i(\mathbf{x}) u_{y,i}(t) \\
\sum_{i=1}^{n_n}\phi_i(\mathbf{x}) u_{z,i}(t) 
\end{array} \right) = \mathbf{B}_\mathbf{u}(\mathbf{x})\mathbf{U}(t).
 \]
\end{linenomath}
where $\mathbf{B}_\mathbf{u}(\mathbf{x})$ is given by:
\begin{linenomath}
 \[\arraycolsep=1.0pt\def\arraystretch{1.0} \mathbf{B}_\mathbf{u}(\mathbf{x}) =   \left( \begin{array}{ccccccccc}
\phi_1(\mathbf{x}) & ... & \phi_{n_n}(\mathbf{x}) & & 0 & & & 0 & \\
 & 0 &  & \phi_1(\mathbf{x}) & ... & \phi_{n_n}(\mathbf{x}) & & 0 & \\
&0& & & 0 & & \phi_1(\mathbf{x}) & ... & \phi_{n_n}(\mathbf{x}) 
\end{array} \right),
 \]
\end{linenomath}
$\mathbf{U}(t)$ is a vector of the DOFs of the displacement in $x$, followed by the $y$-displacement DOFs and then by the $z$-displacement DOFs. The integer $n_n$ is the number of nodes. The basis functions $\phi_i$ are given by the formula:
\begin{linenomath}
\[\arraycolsep=3.0pt\def\arraystretch{1.0} \phi_i(\mathbf{x}) =   \left\{ \begin{array}{clc}
 \frac{(\mathbf{x}-\mathbf{x}_l)(\mathbf{x}_i - \mathbf{x}_l)}{(\mathbf{x}_i-\mathbf{x}_l)(\mathbf{x}_i - \mathbf{x}_l)},& \text{for } \mathbf{x} \in T^{(l)} & l \in \{j_{i,1}, j_{i,2},...,j_{i,n_{\phi_i}}\} \\
  0,                                             & \text{for } \mathbf{x} \in \Omega - \cup_{l \in \{j_{i,1}, j_{i,2},...,j_{i,n_{\phi_i}}\}} T^{(l)}             &     
\end{array} \right. ,
 \]
\end{linenomath}
where $\mathbf{x}_i$ is the position vector of the node $i$ associated with the basis function $\phi_i$ and $x_l$ is the position vector of the normal projection  of $\mathbf{x}_i$ on the plane containing the face opposite to the node $\mathbf{x}_i$ in the tetrahedron $T^{(l)}$. The integer $n_{\phi_i}$ is the number of the tetrahedra that share the node $i$ and the set of indices $\{j_{i,1}, j_{i,2},...,j_{i,n_{\phi_i}}\} \subset \{ 1,2,...,n_e\}$ is the global indices of those tetrahedra, where $n_e$ is the number of tetrahedra in the mesh. The piecewise constant approximation of the pressure can be written as $p(\mathbf{x}, t) \approx \hat{p}(\mathbf{x}, t) = \sum_{j=1}^{n_e} \chi_j(\mathbf{x}) p_j(t) = \mathbf{B_p}(\mathbf{x})^T\mathbf{P}(t)$, $\mathbf{B_p}(\mathbf{x})$ is the vector of the basis functions $\chi_j(\mathbf{x})$, $\mathbf{P}(t)$ is the DOFs of the pressure, and the basis functions $\chi_j(\mathbf{x})$ are given by:
 \begin{linenomath}
\[\arraycolsep=3.0pt\def\arraystretch{1.0} \chi_j(\mathbf{x}) =   \left\{ \begin{array}{cl}
   1,  & \text{for } \mathbf{x} \in T^{(j)} \\
   0,  & \text{for }\mathbf{x} \in \Omega - T^{(j)},                  
\end{array} \right.
 \]
\end{linenomath}
where $T^{(j)}$ is the tetrahedron associated to the $j^\text{th}$ basis function, $\chi_j(\mathbf{x})$. The lowest order RV approximation of the velocity is given by the following expression: $\mathbf{q}(\mathbf{x}, t) = \sum_{k=1}^{n_f}\mathbf{\psi}_k(\mathbf{x})q_k(t) = \mathbf{B}_q(\mathbf{x})^T\mathbf{Q}(t)$, where $n_f$ is the number of faces, $ \mathbf{B}_q(\mathbf{x})$ is the vector of the basis functions, and $\mathbf{Q}(t)$ is the vector of the velocity DOFs.  The vector-valued basis functions $ \mathbf{\psi}_k(\mathbf{x}) $ are given by:
 \begin{linenomath}
 \[\arraycolsep=3.0pt\def\arraystretch{1.0} \mathbf{\psi}_k(\mathbf{x}) =   \left\{ \begin{array}{clc}
 \pm \frac{(\mathbf{x}-\mathbf{x}_k)}{3 |V(T^{(m)})|},   &\text{for }  \mathbf{x} \in T^{(m)} & m \in \{j_{k,1},j_{k,2} \} \\
  0,                                                   &\text{for }  \mathbf{x} \in \Omega - \cup_{m \in \{j_{k,1},j_{k,2} \}} T^{(m)},             &     
\end{array} \right. 
 \]
\end{linenomath}
where $\{j_{k,1},j_{k,2} \} \subset \{1,2,...,n_e\}$ are the indices of the tetrahedrons that share the $k^\text{th}$ face, and $\mathbf{x}_k$ is the position of the node that is not shared by $k^\text{th}$ face in the tetrahedron $T^{(m)}$. The conventional choice of the sign is such that the vector $\mathbf{\psi}_k(\mathbf{x})$ points outward the element $T^{(m)}$ with smallest index $m$. 

We substitute the approximated functions in the weak form~\refp{equ: Weak Form 3 Field a}--\refp{equ: Weak Form 3 Field c} to obtain the discretized system. The resulting linear system of equations that we need to solve each time step is as follows:
\begin{align}
\mathbf{M} \mathbf{P}_k + \alpha \mathbf{D}\mathbf{U}_k  + {\Delta t}_k \mathbf{D}^\mathbf{q}\mathbf{Q}_k =&{\Delta t}_k \mathbf{F}^p_k +\mathbf{M} \mathbf{P}_{k-1} + \alpha \mathbf{D}\mathbf{U}_{k-1} \label{equ: Appendix Discretized 3 Field System a}  \\
- \alpha\mathbf{G}\mathbf{P}_k - \mathbf{E}\mathbf{U}_k   =&-	  \mathbf{F}^\mathbf{u}_k - \mathbf{F}^{\mathbf{u}, bcs}_k \label{equ: Appendix Discretized 3 Field System b}   \\
        {\Delta t}_k \mathbf{G}^p  \mathbf{P}_k -  {\Delta t}_k \mathbf{K} \mathbf{Q}_k =&   {\Delta t}_k \mathbf{F}^{p,bcs}_k, \label{equ: Appendix Discretized 3 Field System c} 
\end{align}
where the subscript $k$ denote that the vector is evaluated (or to be evaluated) at time $t_k$. We combine the equations~\refp{equ: Appendix Discretized 3 Field System a}--\refp{equ: Appendix Discretized 3 Field System c} in one system as follows:
\begin{align}
	\mathbf{L}_k \mathbf{X}_k = \mathbf{F}_{k}  + \mathbf{N}\mathbf{X}_{k-1}, \label{equ: Time Step Forward System }
\end{align}
where
\begin{linenomath}
\[\arraycolsep=3.0pt\def\arraystretch{1.0} \mathbf{L}_k = \left( \begin{array}{ccc}
		\mathbf{M} & \alpha \mathbf{D} & {\Delta t}_k  \mathbf{D}^\mathbf{q}  \\
		-\alpha \mathbf{G} & -\mathbf{E}  & \mathbf{0} \\
                  - {\Delta t}_k  \mathbf{G}^\mathbf{p} &  \mathbf{0} &  {\Delta t}_k \mathbf{K}  \end{array} \right),
\]
\end{linenomath}
\begin{linenomath}
 \[\arraycolsep=3.0pt\def\arraystretch{1.0}\mathbf{N} =  \left( \begin{array}{ccc}
\mathbf{M}& \alpha\mathbf{D}  & \mathbf{0} \\
\mathbf{0} & \mathbf{0}  &  \mathbf{0} \\
\mathbf{0} & \mathbf{0}  &  \mathbf{0} \\
 \end{array} \right), \text{and} \; 
  \mathbf{F}_{k} =  \left( \begin{array}{c}
  {\Delta t}_k \mathbf{F}^p_k\\
 -  \mathbf{F}^\mathbf{u}_k - \mathbf{F}^{\mathbf{u}, bcs}_k  \\
 {-\Delta t}_k \mathbf{F}^{p,bcs}_k \\
 \end{array} \right).
  \]
\end{linenomath}
 \begin{linenomath}
\[\arraycolsep=3.0pt\def\arraystretch{1.0} \mathbf{X}_{k} =  \left( \begin{array}{c}
\mathbf{P_k} \\
\mathbf{U_k} \\
\mathbf{Q_k}
 \end{array} \right).\]
 \end{linenomath}
For derivation purposes, we combine solving for all the $N_t$ time steps, system~\refp{equ: Time Step Forward System }, in one ``global" system (following \citeA{HesseStadler14}) which we write as:
\begin{align}
	\gt{S}\gt{X} = \gt{F}, 
\end{align}
where
\begin{linenomath}
\[\arraycolsep=3.0pt\def\arraystretch{1.0} \gt{S} = \left( \begin{array}{cccccc}
\mathbf{L}_0 & \mathbf{0} & \mathbf{0} & ... & \mathbf{0} & \mathbf{0} \\
-\mathbf{N} & \mathbf{L}_1 & \mathbf{0} & ... & \mathbf{0} & \mathbf{0} \\
:  & :& : & ... & :& :\\
\mathbf{0}   &\mathbf{0}  & \mathbf{0} & ... & \mathbf{L}_{N_t-1} & \mathbf{0} \\
\mathbf{0}   &\mathbf{0}  & \mathbf{0}  & ... & -\mathbf{N} & \mathbf{L}_{N_t}  \end{array} \right),
\gt{X} = \left( \begin{array}{c}
\mathbf{X}_0\\
\mathbf{X}_1\\
:\\
\mathbf{X}_{N_t-1}\\
\mathbf{X}_{N_t} \end{array} \right), \text{and}\;
\gt{F} = \left( \begin{array}{c}
\mathbf{F}_0\\
\mathbf{F}_1\\
:\\
\mathbf{F}_{N_t-1}\\
\mathbf{F}_{N_t} \end{array} \right).\]
\end{linenomath}
 The bar is used in the notation to distinguish the space-time discretized operators and vectors from the space-only discretized operators.

\section{Derivation of Newton Iteration for Estimating the MAP Point}
\label{app: Newton Derivation}

We form the Lagrangian for the constrained optimization problem~\refp{equ: Constrained Minimization} to be:
\begin{align}
 \mathcal{L}(\mathbf{m},\gt{X},\gt{Y}) = 
&\frac{1}{2}(\gt{B}\gt{X} - \mathbf{d}^{\text{obs}}  )^T
\mathbf{\Gamma}_\text{noise}^{-1}( \gt{B}\gt{X}
-\mathbf{d}^{\text{obs}}) \\
+&  \frac{1}{2}(\mathbf{m} - \bar{\mathbf{m}})^T\mathbf{R}
 (\mathbf{m} - \bar{\mathbf{m}}) 
 + \gt{Y}^T(\gt{S}(\mathbf{m})\gt{X}- \gt{F}).   \label{equ: First Lagrangian}
\end{align}
 The adjoint equation, the state equation and the gradient are given by:
 \begin{align}
 \frac{\partial\mathcal{L}}{\partial \gt{X}}(\gt{X}, \gt{Y}, \mathbf{m})   =&
\gt{B}^T\mathbf{\Gamma}_\text{noise}^{-1} ( \gt{B}\gt{X} -
\mathbf{d}^{\text{obs}}  ) + \gt{S}^T(\mathbf{m})\gt{Y} = 0 \quad \text{(adjoint)},
\label{equ: Adjoint Equation}\\
  \frac{\partial\mathcal{L}}{\partial \gt{Y}}(\gt{X}, \gt{Y}, \mathbf{m}) =& \gt{S}(\mathbf{m})\gt{X} -
\gt{F} =
0  \quad \text{(state)},
\label{equ: State Equation}\\
   \frac{\partial\mathcal{L}}{\partial \mathbf{m}}(\gt{X}, \gt{Y}, \mathbf{m}) =& \mathbf{R}(\mathbf{m} -
\bar{\mathbf{m}}) + \gt{C}^T(\mathbf{m})\gt{Y}  \quad \text{(gradient)}.
\label{equ: Gradiant}
\end{align}

 Consider the Lagrangian $\mathcal{L}_\mathcal{H}$ formed as follows:
\begin{align}
 \mathcal{L}_\mathcal{H}:= & {\delta\gt{X}}^T \left(
\gt{B}^T\mathbf{\Gamma}_\text{noise}^{-1} ( \gt{B}\gt{X} - \mathbf{d}^{\text{obs}}  ) 
  + \gt{S}^T(\mathbf{m})\gt{Y}\right)\\
 + &{\delta\gt{Y}}^T \big( \gt{S}(\mathbf{m})\gt{X} - \gt{F}\big)\\
 + &{\delta \mathbf{m}}^T \big( \mathbf{R} (\mathbf{m} - \bar{\mathbf{m}}) + \gt{C}^T(\mathbf{m})\gt{Y}\big),
\end{align}
where $\delta\gt{X}$, $\delta\gt{Y}$ and $\delta\mathbf{m}$ are incremental directions. The incremental forward problem is defined as follows:
\begin{align}
\frac{\partial\mathcal{L}_\mathcal{H}}{\partial \gt{Y}}  &= \gt{S}(\mathbf{m}){\delta\gt{X}} +
\gt{C}(\mathbf{m}){\delta\mathbf{m}} =0.\label{equ: Incremental Forward}
\end{align}
And similarly the incremental adjoint problem is:
\begin{align}
\frac{\partial\mathcal{L}_\mathcal{H}}{\partial \gt{X}} &=
\gt{B}^T\mathbf{\Gamma}_\text{noise}^{-1}\gt{B}  {\delta\gt{X}} + 
\gt{S}^T(\mathbf{m}){\delta\gt{Y}} +\Big( \frac{\partial}{\partial \gt{X}} (\gt{C}^T(\mathbf{m})\gt{Y})\Big)^T\delta \mathbf{m}  \nonumber \\
&=  \mathbf{W}_{\gt{X}\gt{X}}  {\delta\gt{X}} 
 + \gt{S}^T(\mathbf{m}){\delta\gt{Y}} + \mathbf{W}_{\gt{X}\mathbf{m}}(\mathbf{m})\delta \mathbf{m} =
0.\label{equ: Incremental Adjoint}
\end{align}
The $k_\text{th}$ time step in the incremental forward problem will require solving the system:
\begin{align}
\mathbf{L}(\mathbf{m}){\delta\mathbf{X}}^{k} = \mathbf{N}\delta\mathbf{X}^{k-1} - \mathbf{C}^{k}(\mathbf{m})\delta\mathbf{m},  
\end{align}
and the $k_\text{th}$ time step in the incremental adjoint problem requires solving the following system:
\begin{align}
\mathbf{L}^T(\mathbf{m}) {\delta \mathbf{Y}}^{k} = \mathbf{N}^T{\delta\mathbf{Y}}^{k+1} -
 \mathbf{W}_{\mathbf{X}\mathbf{X}} {\delta\mathbf{X}}^{k} -\mathbf{W}^{k}_{\mathbf{X}\mathbf{m}}(\mathbf{m})\delta \mathbf{m}. 
\end{align}
The Hessian action on $\delta\mathbf{m}$ is given by:
\begin{align}
  \frac{\partial\mathcal{L}_\mathcal{H}}{\partial \mathbf{m}}
&=   \frac{\partial}{\partial \mathbf{m}} ( {\delta\gt{X}}^T\gt{S}^T(\mathbf{m})\gt{Y})
+  \frac{\partial}{\partial \mathbf{m}} ( {\delta\gt{Y}}^T\gt{S}(\mathbf{m})\gt{X})
+    \mathbf{R}\delta\mathbf{m} + \frac{\partial}{\partial \mathbf{m}}  ({\delta
\mathbf{m}}^T \gt{C}^T(\mathbf{m})\gt{Y})\nonumber\\
&=  \mathbf{W}_{\mathbf{m}\gt{X}}(\mathbf{m}) \delta\gt{X} +  \gt{C}^T(\mathbf{m}) \delta\gt{Y}
+  \mathbf{R}\delta\mathbf{m} +
\mathbf{R}_{\mathbf{m}\mathbf{m}}(\mathbf{m})\delta\mathbf{m}. \label{equ: Hessian Expression} 
\end{align}
Substituting the values of $\delta\mathbf{X}$ and $\delta\mathbf{X}$ from 
the systems~\refp{equ: Incremental Forward} and \refp{equ: Incremental Adjoint} into the expression~\refp{equ: Hessian Expression} gives the Hessian expression:
\begin{align}
 \mathbf{H} = &
 \mathbf{R} + \mathbf{R}_{\mathbf{m} \mathbf{m}}(\mathbf{m}) +\gt{C}^{T}(\mathbf{m}) \gt{S}^{-T}(\mathbf{m}) (
\mathbf{W}_{\gt{X}\gt{X}} \gt{S}^{-1}(\mathbf{m}) \gt{C}(\mathbf{m})-   \mathbf{W}_{\gt{X}\mathbf{m}}(\mathbf{m}) ) \nonumber\\
 &-
\mathbf{W}_{\mathbf{m} \gt{X}}(\mathbf{m})\gt{S}^{-1}(\mathbf{m})\gt{C}(\mathbf{m}).
\end{align}

\color{black}

\FloatBarrier
\acknowledgments
This work was supported by National Science Foundation (NSF) Grants CBET–1508713 and ACI-1550593, and DOE grant DE-SC0019303. A.A. would like to acknowledge funding from the Ministry of Education in Saudi Arabia. 
The authors would like to thank Dr.\ Thomas Burbey for providing the Nevada experiment
GPS and well head data and for his helpful comments.
The authors would also like to thank Dr.\ Umberto Villa, a main developer of
 \texttt{hIPPYlib}, for the many helpful discussions during the course of this
research,
 and Dr.\ Jeonghun (John) Lee for his help with the convergence verification
 of the 3D 3-field poroelasticity forward solver to a manufactured solution. 
The Envisat SAR imagery
used in this study can be downloaded through the UNAVCO Data Center SAR archive.


%
%

\end{document}